\documentclass[twocolumn]{aastex62}
\usepackage{amsmath}
\usepackage{amssymb}
\usepackage{mathtools}
\usepackage{multirow}

\providecommand{\given}{\ensuremath{\hspace{0.05em}\mid\hspace{0.05em}}}

\providecommand{\given}{\ensuremath{\hspace{0.01em}\mid\hspace{0.01em}}}

\providecommand{\gaia}{Gaia}
\providecommand{\gdr}[1]{GDR{#1}}
\providecommand{\gedr}[1]{GeDR{#1}}

\providecommand{\release}{GDR3}

\providecommand{\parallax}{\ensuremath{\varpi}}

\providecommand{\propm}{\ensuremath{\mu}}
\providecommand{\pmra}{\ensuremath{\mu_{\alpha*}}}
\providecommand{\pmdec}{\ensuremath{\mu_{\delta}}}
\providecommand{\sigparallax}{\ensuremath{\sigma_{\varpi}}}
\providecommand{\cov}{\ensuremath{\Sigma}}
\providecommand{\glon}{\ensuremath{l}}
\providecommand{\glat}{\ensuremath{b}}
\providecommand{\hp}{\ensuremath{p}} 
\providecommand{\dist}{\ensuremath{r}}

\providecommand{\vra}{\ensuremath{v_{\alpha*}}}
\providecommand{\vdec}{\ensuremath{v_{\delta}}}

\providecommand{\degree}{\ensuremath{^\circ}}
\providecommand{\kms}{\ensuremath{\mathrm{km}\,\mathrm{s}^{-1}}}
\providecommand{\pc}{\ensuremath{\mathrm{pc}}}
\providecommand{\kpc}{\ensuremath{\mathrm{kpc}}}

\providecommand{\maspyr}{\ensuremath{\mathrm{mas\,\mathrm{yr}^{-1}}}}

\received{2023-07-27}
\revised{2023-10-08 and 2023-10-31}
\accepted{2023-10-31}
\journalinfo{Accepted to AJ}

\shorttitle{Kinegeometric distances and velocities in \gaia\ DR3}
\shortauthors{C.A.L.\ Bailer-Jones}

\begin{document}



\title{Estimating distances from parallaxes. VI:\\
A method for inferring distances and transverse velocities from parallaxes and proper motions\\demonstrated on Gaia Data Release 3
}


\author{C.A.L.\ Bailer-Jones}
\affil{Max Planck Institute for Astronomy, Heidelberg, Germany}

\begin{abstract}
  
The accuracy of stellar distances inferred purely from parallaxes degrades rapidly with distance.
Proper motion measurements, when combined with some idea of typical velocities, provide independent information on stellar distances.
  Here I build a direction- and distance-dependent model of the distribution of stellar velocities in the Galaxy, then use this together with parallaxes and proper motions to infer kinegeometric distances and transverse velocities for stars in \gaia\ DR3.
  Using noisy simulations I assess the performance of the method and compare its accuracy to purely parallax-based (geometric) distances. Over the whole \gaia\ catalogue, kinegeometric distances are on average 1.25 times
  more accurate than geometric ones. This average masks a large variation in the relative performance, however. Kinegeometric distances are considerably better than geometric ones
  beyond several \kpc, for example.
    On average, kinegeometric distances can be measured to an accuracy of 19\% and velocities
($\sqrt{\vra^2 + \vdec^2}$)  to 16\,\kms\
  (median absolute deviations).
In \gaia\ DR3, kinegeometric distances are smaller than geometric ones on average for distant stars, but the pattern is more complex in the bulge and disk.
With the much more accurate proper motions expected in \gaia\ DR5, a further improvement in the distance accuracy by a factor of (only) 1.35 on average is predicted 
(with kinegeometric distances still 1.25 times more accurate than geometric ones).
The improvement attained from proper motions is limited by the width of the velocity prior, in a way that the improvement from better parallaxes is not limited by the width of the distance prior.
\end{abstract}

\keywords{catalogs -- Bayesian statistics -- distance indicators -- stellar distances -- stellar motion -- astrometry -- parallax -- proper motions}


\section{Introduction}\label{sec:introduction}

Precise parallaxes may be simply inverted to get precise distance estimates of individual stars. This is not possible for low precision parallaxes, however, because the inversion induces a disproportionately large random error and bias \citep{2015PASP..127..994B, 2018A&A...616A...9L}.
Of the 1.47 billion stars with parallaxes in \gaia\ Data Release 3 (\gdr3) \citep{2023A&A...674A...1G}, 87\% have a parallax SNRs less than five, and 78\% less than three.
Even in the final \gaia\ data release the parallax SNRs will only improve by a factor of two for most stars,
leaving 75\% still with parallax SNRs less than five (60\% less than three).

A more considered approach than parallax inversion is therefore required for distance estimation. This has been framed as a probabilistic inference (Bayesian) problem in the present series of papers.
In the first paper \citep{2015PASP..127..994B} I looked at the general problem of inferring distances from parallaxes, and introduced the exponentially-decreasing space density (EDSD) prior. The second paper \citep{2016ApJ...832..137A} introduced a more sophisticated distance prior based on a model of the Milky Way. Both this and simpler isotropic priors were applied in the third paper \citep{2016ApJ...833..119A} to estimate distances for all two million stars in \gdr1\ that had parallaxes. Paper four \citep{2018AJ....156...58B} built a direction-dependent EDSD prior for the whole Galaxy using a mock \gaia\ catalogue and used this to infer geometric distances for all 1.33 billion stars in \gdr2\ with parallaxes. \cite{2021AJ....161..147B}, the fifth in the series, did this again for the higher precision parallaxes in \gedr3\ using a more flexible direction-dependent distance prior (the generalized gamma distribution). This paper also extended the method to take advantage of the \gaia\ colour and apparent magnitude of each star which, together with a model for the colour--absolute magnitude diagram (which also varied over the sky), gives information on distance (as already explored in \citealt{2016ApJ...832..137A}). These distances, which we called {\em photogeometric distances}, were published for 1.35 billion stars, along with geometric distances for 1.47 billion stars.

Several other authors have published distance estimation methods or catalogues using the \gaia\ parallaxes. Some focus on specific objects, such as Wolf-Rayet stars \citep{2020MNRAS.493.1512R} or Mira variables \citep{2023MNRAS.523.2369S} or stellar clusters \citep{2020A&A...640A...1C, 2020A&A...644A...7O}.
Others have performed large scale distance estimation of broad types of individual stars for which more general priors must be used.
Both \cite{2022A&A...662A.125F} and 
\cite{2022A&A...658A..91A} combined \gaia\ parallaxes with optical and infrared photometry from multiple surveys
to infer intrinsic stellar properties, line-of-sight extinctions, and distances for over a hundred million stars.
\cite{2023A&A...674A..27A} used the \gaia\ BP/RP spectra and parallaxes to also estimate stellar parameters, extinctions, and distances, and published these as part of \gdr3\ for half a billion stars.
Using the publicly-released subset of the \gdr3\ BP/RP spectra, \cite{2023MNRAS.524.1855Z} also estimated stellar parameters (albeit parallax rather than distance, so the inversion problem remains).
Although conceptually similar to the approach used to produce photogeometric distances in~\cite{2021AJ....161..147B}, these large-scale distance estimation projects differ in their methods, and by using the spectral energy distribution make stronger assumptions about the stars themselves.

Stellar proper motions can provide further distance information.  A star moving with an angular velocity \propm\ at distance \dist\ has a transverse velocity of
\begin{alignat}{2}
  \frac{v}{\kms} \,=&\, k \ \frac{\propm}{\maspyr} \frac{\dist}{\kpc} \hspace*{0.5em}\textrm{with}\hspace*{0.5em}k=4.740471 \ .
\label{eqn:transvel}
\end{alignat}
Assuming stars have a limited range of plausible velocities, a measurement of the proper motion puts a constraint on the plausible distance. A velocity prior quantifies what we mean by ``plausible'' and inference delivers a posterior over distance.
Alone this would provide a rather poor distance estimate, but when combined with a parallax measurement we achieve a {\em kinegeometric distance} estimate that is potentially more accurate than a purely geometric one. This is of particular interest for distant stars in \gaia, as these tend to have very low SNR parallaxes but still relatively accurate proper motions. I explore this approach in this paper. 

The use of proper motions to aid distance estimation is not new \citep[see][section 2.2 for an overview]{1998gaas.book.....B}.
A simple case is to adopt a model of circular rotation in the Galaxy (a rather narrow prior), and to estimate the distance directly from the proper motion
  (e.g.\ \citealt{2022AJ....164..133R}).
Proper motions have long been used to estimate the mean distance to a group of stars,
often under the name of statistical parallaxes, by measuring the radial velocities as well as the proper motions (e.g. \citealt{1936ApJ....84..555S}, \citealt{1975AJ.....80..199H}).
\cite{2019MNRAS.487.3568S} adopted a related approach to reduce the bias in parallax-based distance estimates, and used this to estimate distances to 7 million stars with RVs in \gdr2.
\cite{2021A&A...650A.112Z} used a similar approach to the one in the present paper to estimate distances to OBA stars in the Galactic disk from \gedr3. As the target population was narrowly defined, the distance and velocity priors could be tailored to the expected distribution for young stars out to 3--4\,\kpc.

In the present paper I use a direction and distant-dependent velocity prior derived from a Galaxy model to estimate kinegeometric distances to a random subset of stars from the entire \gdr3\ catalogue \citep{gdr3_from_esa}.
I compare the results to geometric distances, and assess the performance of both types of estimate using a mock catalogue.
The kinegeometric method also provides estimates of the two transverse velocity components.
Radial velocities are not used: When I refer to ``velocity'' in this paper it means the total transverse velocity, not the 3D velocity.
In contrast to the previous three papers in this series, I do not publish a catalogue because, as we shall see, the improvements from using proper motions are substantial only in a limited part of parameter space.

\vspace*{1em}
\section{Method}\label{sec:method}

The prior is constructed for each HEALpixel \citep{2005ApJ...622..759G} independently.
As in \cite{2021AJ....161..147B}, I
use the equatorial, nested scheme
at level 5, giving 12\,288 HEALpixels with an area of 3.36 square degrees each.
All sky plots in this paper are a Mollweide equal-area projection in Galactic coordinates, with north up, east increasing to the left, and $(\glon, \glat) = (0\degree, 0\degree)$ in the centre. They were plotted using
{\em healpy} \citep{2019JOSS....4.1298Z} and show grid lines at 45\degree\ in latitude and longitude.
All other plots, as well as the processing and analysis, were done in R (\url{r-project.org}).

\subsection{Distance and transverse velocity posterior}\label{sec:method_posterior}

\cite{2021AJ....161..147B} estimated geometric distances using a prior that varied only with distance ($\dist$) for each HEALpixel (\hp).
The product of this prior and the likelihood is the unnormalized posterior  probability distribution function (PDF)
\begin{equation}
  P_{\rm g}^*(\dist \given \parallax, \sigparallax, \hp) \,=\  P(\parallax \given \dist, \sigparallax) \, P(\dist \given \hp) \ .
  \label{eqn:distpost_geo}
\end{equation}
The likelihood is a Gaussian in parallax (\parallax) with uncertainty \sigparallax, as defined in section 2.2 of \cite{2021AJ....161..147B}.
For comparison purposes I below report geometric distances inferred from this posterior. These use the same input data, parallax zeropoint, and prior as in \cite{2021AJ....161..147B}, so are statistically identical.

Here I introduce the {\em kinegeometric distance}, which uses the proper motions ($\pmra, \pmdec$) in addition to the parallax.
Proper motions are connected to the distance via equation~\ref{eqn:transvel}, so to exploit them we must introduce both a prior on the velocities ($\vra, \vdec$) and a likelihood on the proper motions. Combining all this we get the 3D kinegeometric posterior PDF over distance and velocity
\begin{alignat}{2}
  & P_{\rm kg}^*(\dist, \vra, \vdec \given \parallax, \pmra, \pmdec, \cov, \hp) \,=\ \nonumber \\
  & \hspace{1em} \underbrace{P(\parallax, \pmra, \pmdec \given \dist, \vra, \vdec, \cov)}_{\text{likelihood}} \ \underbrace{P(\dist, \vra,\vdec \given \hp)}_{\text{prior}} \ ,
  \label{eqn:kinegeopost}
\end{alignat}
the $^*$ symbol indicating that the PDF is unnormalized.
The likelihood is a 3D Gaussian with mean
\begin{alignat}{2}
\left( \frac{1}{\dist}, \frac{\vra}{k\dist}, \frac{\vdec}{k\dist} \right)
\end{alignat}
and covariance \cov, which is the $3\times 3$ variance--covariance matrix of the parallax and proper motions published in \gdr3\ for each star \citep{lindegren_etal_gedr3_astrometry,2023A&A...674A...1G}.
The form of the likelihood shows that the more precise the measured proper motion, the more precise the estimated value of $v/\dist$.
As with the geometric distances, I use the parallax zeropoint of \cite{lindegren_etal_gedr3_parallaxzp}. I make no modifications to the proper motions.
The prior will be introduced below.
From the 3D posterior we can estimate the distance and the transverse velocity components by marginalization.
As I numerically sample the posterior (section~\ref{sec:sampling}), this marginalization is trivial.

The main goal of this work is to assess the improvement in distance estimates by using the proper motions in addition to the parallaxes, but I will also investigate the velocities estimated by equation~\ref{eqn:kinegeopost}. Transverse velocities can alternatively be obtained by multiplying the proper motions by the geometric distance.
These I refer to as {\em geometric} velocities, and will compute them for comparison purposes.
I do not consider estimating velocities by dividing the proper motion by the parallax: this is very biased, and fails completely when the parallax is not positive, which is the case for 24\% of sources with parallaxes in \gdr3.

\subsection{Velocity prior}\label{sec:velocity_prior}

\begin{figure}
\begin{center}
  \includegraphics[width=0.49\textwidth, angle=0]{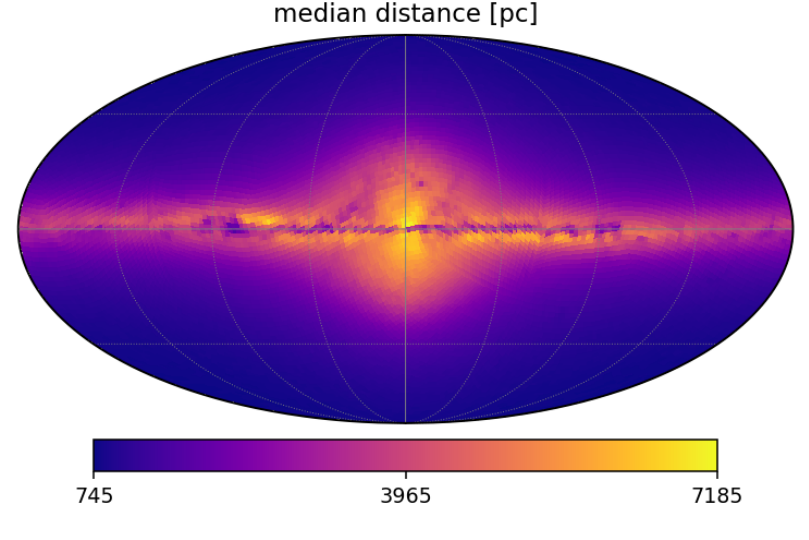}
\caption{Median of the distance prior in each HEALpixel, computed from the mock catalogue of \cite{2020PASP..132g4501R}.
\label{fig:distance_prior_median_skyplot}}
\end{center}
\end{figure}

The distance--velocity prior in equation~\ref{eqn:kinegeopost} can be decomposed without loss of generality as
\begin{alignat}{2}
  P(\dist, \vra,\vdec \given \hp) \,&=\  P(\vra, \vdec \given \dist, \hp) \ P(\dist \given \hp) \ .
  \label{eqn:velocity_prior}
\end{alignat}
For the second term -- the distance prior -- I adopt the same one used for the geometric distances. This is described in \cite[section 2.3]{2021AJ....161..147B}. It was built from the mock \gaia\ catalogue of \cite{2020PASP..132g4501R}
and is summarized in Figure~\ref{fig:distance_prior_median_skyplot}.
The first term in equation~\ref{eqn:velocity_prior}, the velocity prior, I compute here from the same mock catalogue, which is based on the Galaxy model of \cite{2003A&A...409..523R} plus some later modifications.
This model comprises several Galactic populations -- bulge, thin and thick disks of a range of ages, stellar halo, dark matter halo, interstellar medium -- with velocities computed self-consistently with the mass distribution via age--velocity relations.
The disk model includes a warp and flare and the outer bulge is strongly oblate (a bar). I exclude all star clusters.

Stellar velocity distributions vary considerably with direction and distance, so a dependence on both is retained.  The prior should be a continuous -- and ideally smooth --function of distance and direction. Here I compute it independently for each HEALpixel (direction) so it is not smooth across HEALpixel boundaries, but it is smooth in distance.
For a given HEALpixel, I investigated how the velocity distribution varied over distance shells, a shell being defined as the volume between two spheres of different radii centered on the Sun. With sufficient stars, the 2D distribution in each shell could be approximated well with a 2D Gaussian.

\begin{figure*}
\begin{center}
  \includegraphics[width=0.99\textwidth, angle=0]{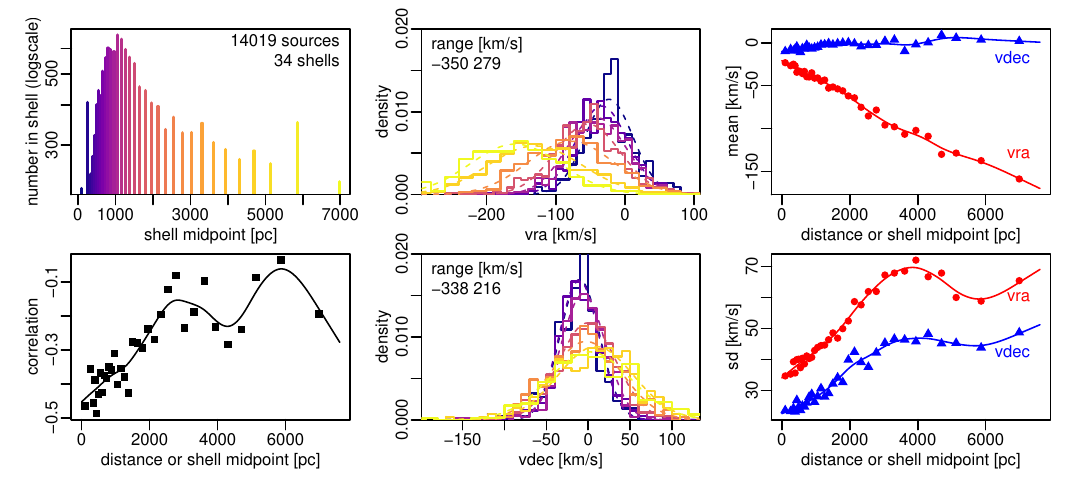}
  \caption{Velocity prior for HEALpixel 6200 at $(\glon, \glat) = (285.7\degree, 34.8\degree)$.
    The top left panel shows the positions of the mid points of the shells and the number of sources in them (on a log scale).
    The two central panels show the distribution of the velocities (top \vra, bottom \vdec) for some selected shells, with the colours denoting the shell as in the top left panel.
    The dashed lines show the fitted Gaussians.
    The two right panels show the values of the fitted means (upper) and fitted standard deviations (lower) for all shells, as red circles for \vra\ and as blue triangles for \vdec. The lines show the spline fits, which extend to the outer radius of the last shell, here 7597\,\pc.
    The bottom left panel shows the fitted correlations of the Gaussians in each shell along with its spline fit, but this is not used (the correlation is set to zero in the model).
\label{fig:velocity_prior_06200}}
\end{center}
\end{figure*}

The task of building a prior then falls to determining how the mean and variance--covariance of this2D Gaussian should vary with distance.
Figure~\ref{fig:velocity_prior_06200} supports the following explanation.
I first construct a series of contiguous shells moving outwards from the Sun, defined such that the volume of each shell is a fixed multiple $\lambda$ of the volume of the previous shell. When the outer radius of the first shell is fixed (its inner radius is zero), this recurrence relation defines all shells.  If stars were uniformly distributed in volume, then $\lambda=1$ would produce shells with the same number of stars.

After experimenting to achieve a trade-off between having sufficiently small shells to capture the variation of the velocity distribution, yet sufficiently large shells to allow an accurate Gaussian fit, I selected $\lambda=1.3$ and an initial outer shell radius of 200\,\pc. I generate a set of 45 shells such that the outer radius of the last shell is 15.3\,\kpc. Stars beyond this do not contribute to the prior. \gaia\ astrometry on more distant stars is poor, so larger distances are not worth accommodating.
Some HEALpixels at high Galactic latitudes still have too few stars, so I merge successive shells (starting from the Sun) to attain at least 200 stars per shell. At higher latitudes this leads to the outermost shell having an outer radius less than 15.3\,\kpc\ (as there are fewer than 200 stars beyond).
The smallest outermost radius is 3.2\,\kpc\ with a median of 9.0\,\kpc.
Among the 12\,228 HEALpixels, the number of shells varies from 11 to 45 with a median value of 36.
The number of stars per HEALpixel varies from 3200 to 3.3 million with a median of 18 thousand.  
The top-left panel of Figure~\ref{fig:velocity_prior_06200} shows an example of the shell structure for one HEALpixel.

Once the shells have been fixed, I compute robust estimates\footnote{I use the median and half the difference of the 84th and 16th percentiles of the distribution.} of the mean and standard deviation of the two velocities, as well as their correlation coefficient, for the stars in each shell.
I then fit a smoothing spline to each of the parameters independently as a function of
distance (right panels of Figure~\ref{fig:velocity_prior_06200}). The number of degrees of freedom of these fits is an increasing function of the number of shells, varying from 3 to 10 with a median of 8. Beyond the outermost radius the Gaussian parameters are kept constant at the value at the outermost radius.
As the fit for the standard deviations could go to very small or even negative values, I force the fit of the two standard deviations never to drop below 2\,\kms.

\begin{figure*}
\begin{center}
  \includegraphics[width=0.99\textwidth, angle=0]{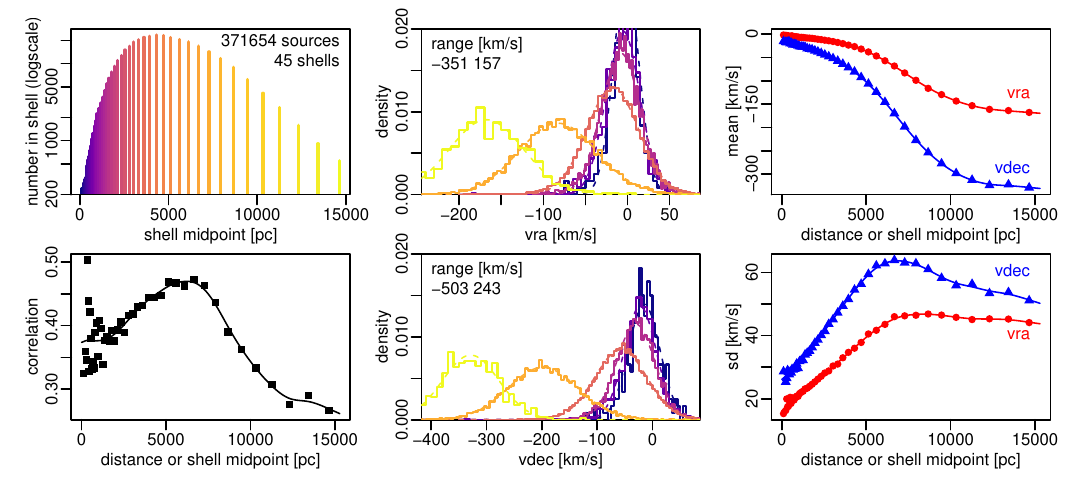}
  \caption{As Figure~\ref{fig:velocity_prior_06200} but now for HEALpixel 7593 at
HEALpixel 7593 at $(\glon, \glat) = (29.0\degree, 7.7\degree)$. 
\label{fig:velocity_prior_07593}}
\end{center}
\end{figure*}

The results of this process for a relatively sparse field at high latitude, where some shells are merged, are shown in Figure~\ref{fig:velocity_prior_06200}.
In this particular example the variation of the correlation with distance is overfit; in other HEALpixels there is no clear correlation.
For this reason I do not use these fits and set the correlation to be zero for all distances for all HEALpixels; a conservative choice, making the prior less informative than it otherwise would be.
Figure~\ref{fig:velocity_prior_07593} summarizes the prior for a HEALpixel at a lower latitude towards the Galactic bulge, with many more stars where no shells are merged. Note the large change in the mean velocities for the most distant shells.

When building this velocity prior I excluded stars from the mock catalogue that are fainter than the expected \gaia\ magnitude limit for that HEALpixel, as was done for the distance prior in \cite{2021AJ....161..147B} (see section 1 of that reference).

I experimented with making the velocity prior a function of magnitude or colour. For magnitude this accounted for little additional variance. Introducing a dependence on colour was much harder, because some HEALpixels show gaps in their colour distribution, which would necessitate additional modelling assumptions to get a continuous prior. 

\begin{figure*}
\begin{center}
  \includegraphics[width=0.49\textwidth, angle=0]{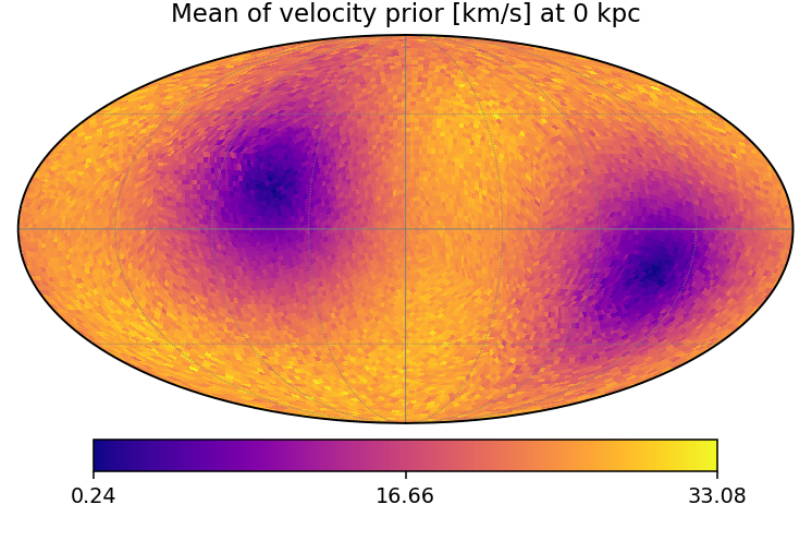}
  \includegraphics[width=0.49\textwidth, angle=0]{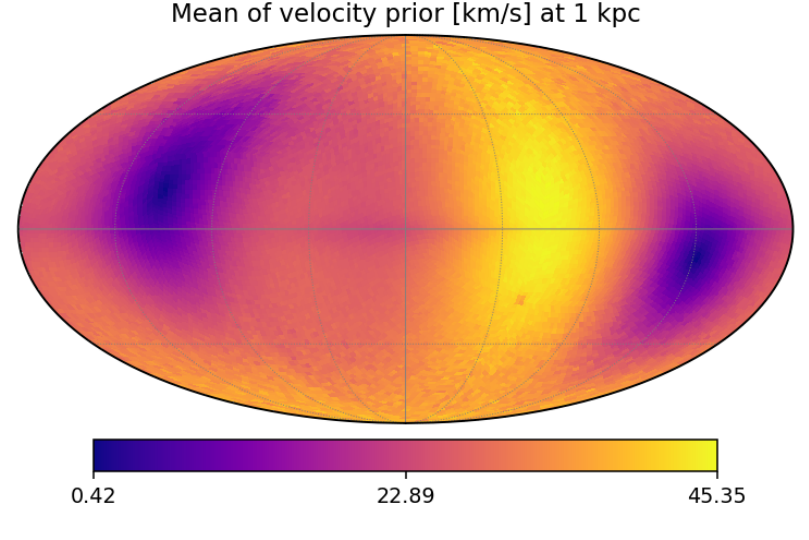}
  \includegraphics[width=0.49\textwidth, angle=0]{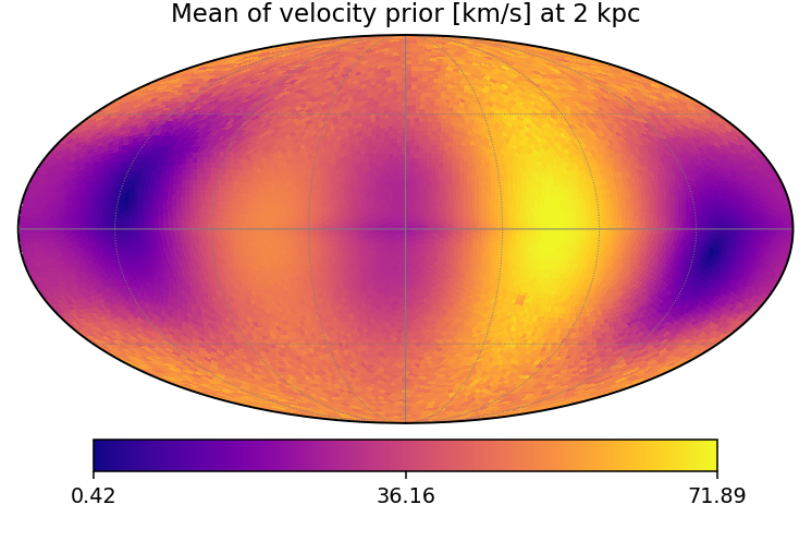}
  \includegraphics[width=0.49\textwidth, angle=0]{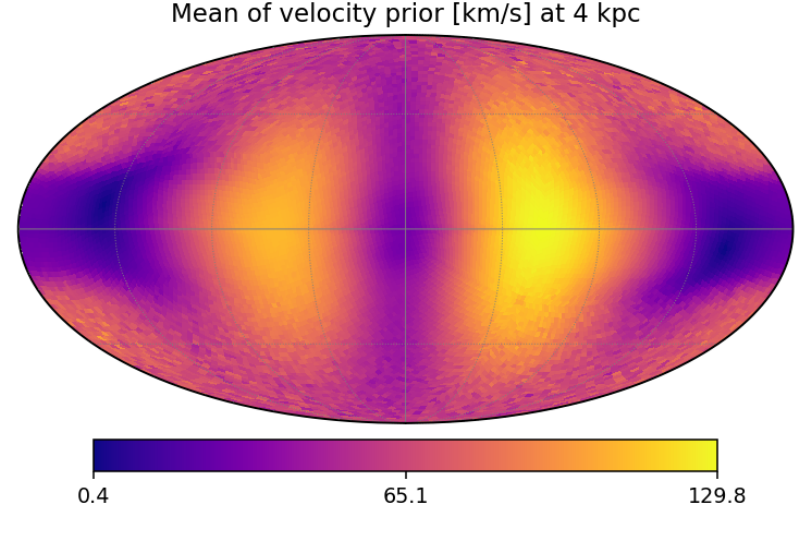}
  \includegraphics[width=0.49\textwidth, angle=0]{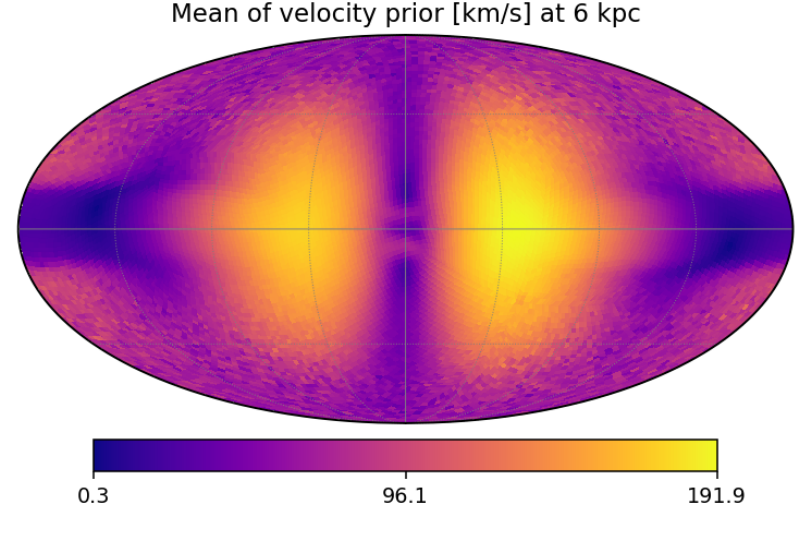}
  \includegraphics[width=0.49\textwidth, angle=0]{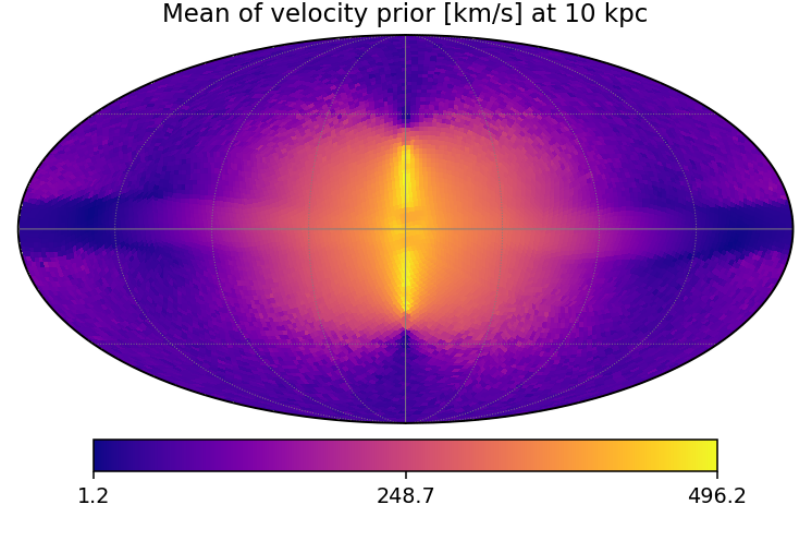}
  \caption{Variation of the mean of the velocity prior over the Galactic sky at six different distances. The quantity shown
    is $\sqrt{\vra^2 + \vdec^2}$, where \vra\ and \vdec\ are the means of the two components of the velocity prior at the specified distance in each HEALpixel. In each distance slice the colour bar covers the full range of velocities plotted, and is different for each slice.
\label{fig:veltotmeanprior_skyplot}}
\end{center}
\end{figure*}

\begin{figure*}
\begin{center}
  \includegraphics[width=0.49\textwidth, angle=0]{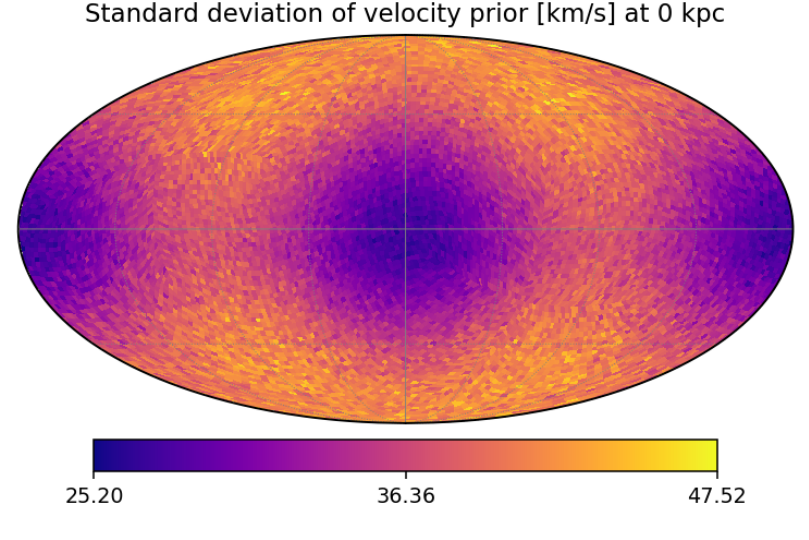}
  \includegraphics[width=0.49\textwidth, angle=0]{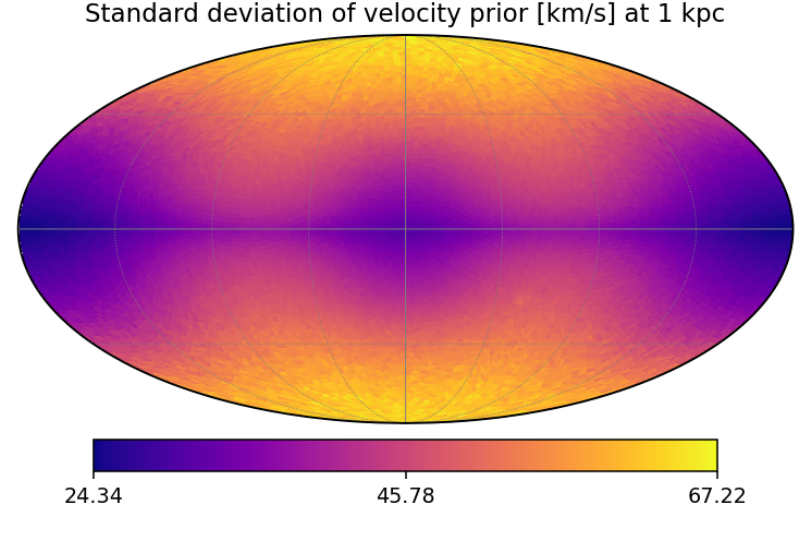}
  \includegraphics[width=0.49\textwidth, angle=0]{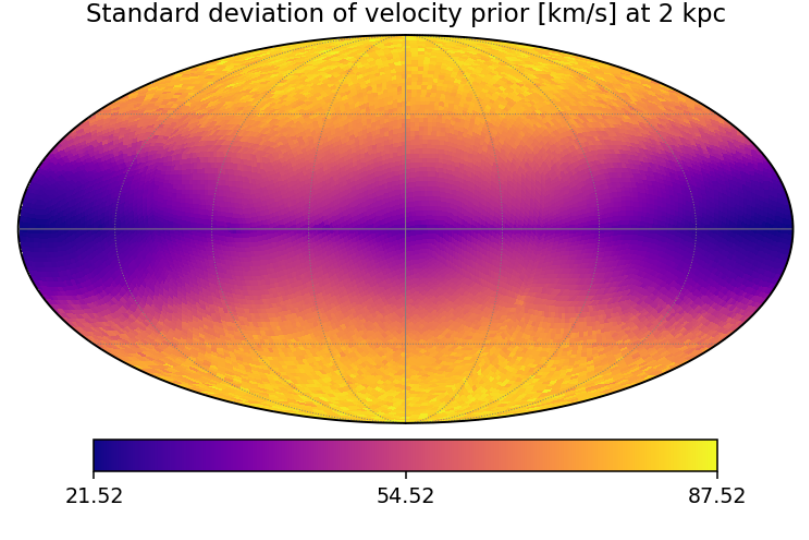}
  \includegraphics[width=0.49\textwidth, angle=0]{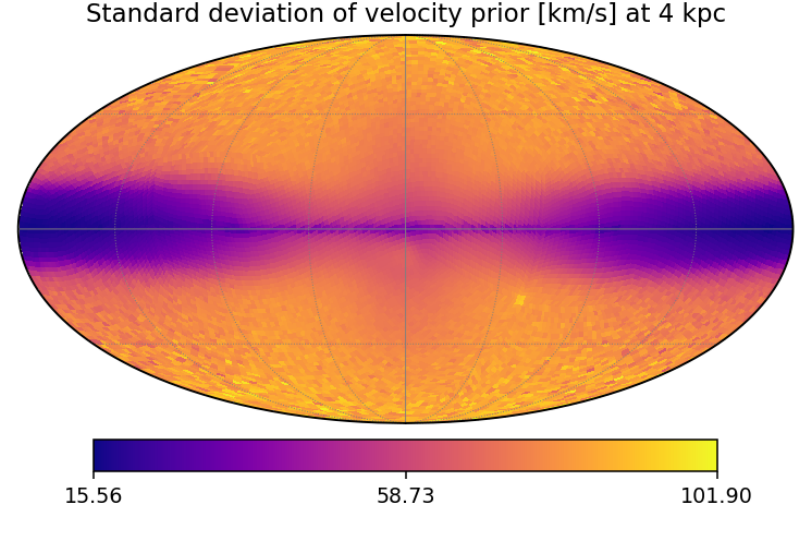}
  \includegraphics[width=0.49\textwidth, angle=0]{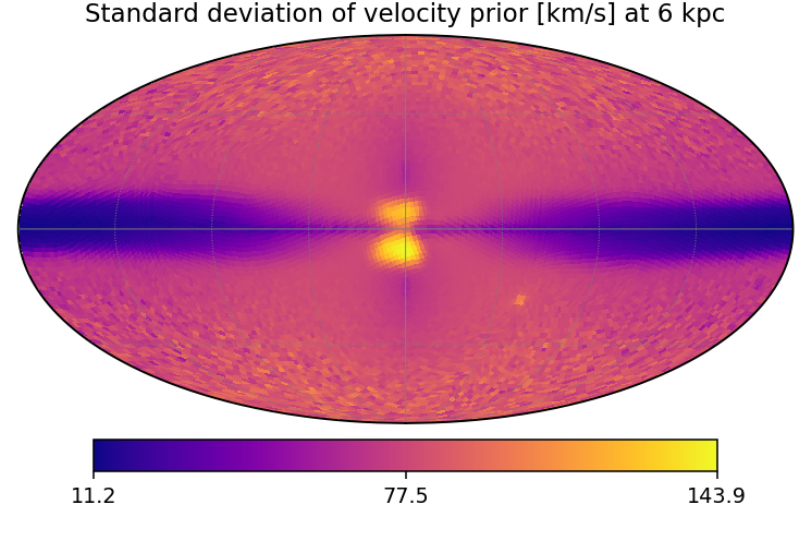}
  \includegraphics[width=0.49\textwidth, angle=0]{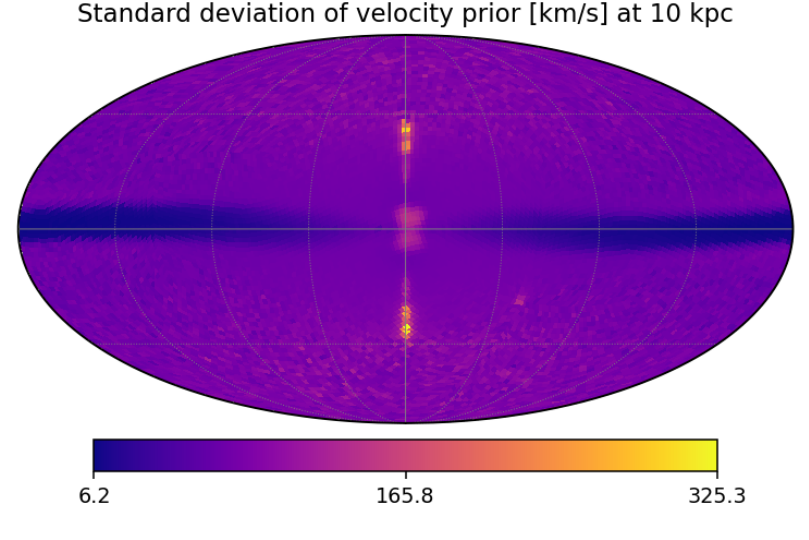}
  \caption{As Figure~\ref{fig:veltotmeanprior_skyplot}, but now showing $\sqrt{\sigma_{\alpha*}^2 + \sigma_{\delta}^2}$,
  where $\sigma_{\alpha*}$ and $\sigma_{\delta}$ are the standard deviations of the two components of the velocity prior.
\label{fig:veltotsdprior_skyplot}}
\end{center}
\end{figure*}

Figure~\ref{fig:veltotmeanprior_skyplot} shows how the magnitude of the mean of the velocity prior varies over
the sky for several distance slices. The velocities of nearby stars are small, but increase at larger distances as we probe the faster orbits in the central regions of the Galaxy (note the change in colour scale across the slices).
The low velocity regions near $(\glon, \glat) = (64\degree, 16\degree)$ and $(\glon, \glat) = (244\degree, -16\degree)$ in the 0\,\kpc\ distance slice are the approximate solar apex and antapex directions (respectively), the direction in which the Sun is moving through the Galaxy.

The inference (equation~\ref{eqn:kinegeopost}) uses the prior on the two components of the velocity, not the total transverse velocity.  
Plots like Figure~\ref{fig:veltotmeanprior_skyplot} for the individual components are less astrophysically informative, however, because they show features of the (equatorial) coordinate system used. An artefact of this is nonetheless visible at the south equatorial pole
$(\glon, \glat) = (302.9\degree, -27.1\degree)$ in some of the distance slices in Figure~\ref{fig:veltotmeanprior_skyplot}.
This is a projection effect of the velocity vector onto the rapidly changing \vra\ and \vdec\ basis vectors at the poles.
In principle we should not see it in the total transverse velocity, but because the prior is computed over the finite area of the HEALpixel, the average shows large jumps.
The same effect occurs at the north equatorial pole, $(\glon, \glat) = (122.9\degree, +27.1\degree)$, but is barely visible.
Using a smaller HEALpixel would mitigate this effect, but create others, and is a limitation of a discrete prior.

Figure~\ref{fig:veltotsdprior_skyplot} shows how the square root of the sum of squares of the standard deviations of the velocity prior varies. This too is not directly used in the inference, but it gives an idea of the width of the velocity prior, which influences how precisely distance can be constrained for a given proper motion (discussed in appendix~\ref{sec:how_it_works}). The velocity constraint tends to be stronger at lower latitudes, as there is less dispersion about the mean rotational velocity of disk stars at a given distance and longitude. At higher latitudes the dispersion increases significantly because the more isotropic orbits of halo stars dominate the tails of the samples. While not immediately visible due to the variable colour scale, the velocity constraint is generally weaker (larger standard deviation) at larger distances.

\newpage
\subsection{Posterior sampling}\label{sec:sampling}

I sample the 3D posterior (equation~\ref{eqn:kinegeopost}) for each star with an MCMC method \citep{2013PASP..125..306F}.\footnote{As the posterior is Gaussian in velocity for a given distance, one could instead sample over distance and then compute the marginal posteriors for the velocity.}
The distance is generally strongly correlated with the velocity components, because for a given proper motion an increase in the distance can be compensated for by a decrease in the velocity. Such correlations did not pose a problem for the sampler, and estimating instead the true angular velocities -- which would not be correlated -- did not confer particular advantages.
After some experimentation I found that 60 burn-in iterations followed by 200 sampling iterations with 20 walkers gave good convergence. 
The acceptance rate was around 0.6, and 
the sample chains were thinned before parameters were estimated.
For each parameter I estimate the 16th, 50th, and 84th percentiles, the median providing the parameter estimate and the other two the lower and upper 1$\sigma$-like uncertainty estimates.


\vspace*{1em}
\section{Performance on the mock \gdr3\ catalogue}\label{sec:results_mock}


To estimate the accuracy of the distance and velocity estimates, I apply the method to a noisy version of the mock \gaia\ catalogue. This is achieved by adding Gaussian random noise to the parallaxes and proper motions at the expected noise level in \gdr3\ \citep{2020PASP..132g4501R}.
The noise-free version of this same catalogue was used to build the priors, so this assessment shows what the method can achieve before being affected by any mismatch between the prior and reality.



I analyse the results over the whole sky using two samples.
The {\em constant fraction sample} selects 0.8\% of the stars at random from each HEALpixel to give a total of 10.3 million sources.
Like the full \gaia\ sample, this is dominated by faint and low-accuracy parallax stars, many of which lie in the Galactic plane and bulge. While this is useful for examining trends in performance with distance and magnitude, it is not as useful for the examining the variation over the sky. For that purpose I use the
{\em constant number sample}, which comprises approximately 900 sources selected at random per HEALpixel, for a total of about 11.5 million sources.
Both samples include only stars that are brighter than the prior magnitude limit in each HEALpixel (section~\ref{sec:velocity_prior}).

\subsection{Overall performance}

\begin{figure*}
\begin{center}
  \includegraphics[width=0.75\textwidth, angle=0]{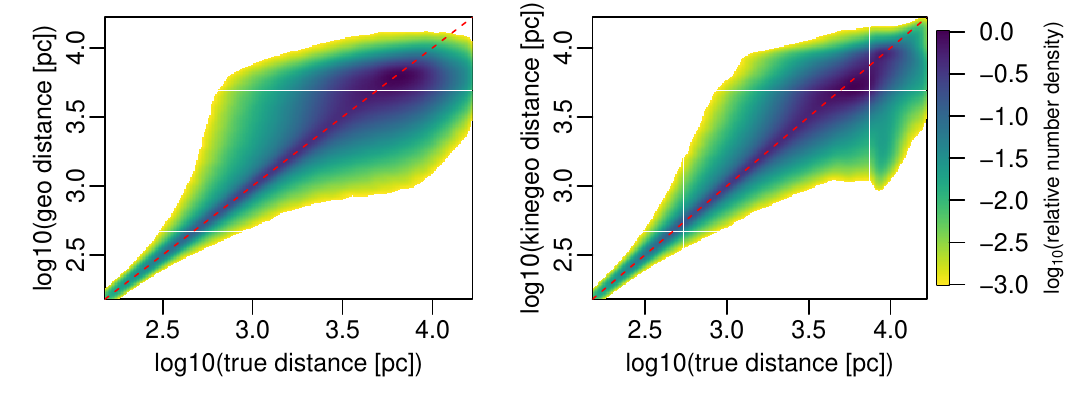}
  \includegraphics[width=0.75\textwidth, angle=0]{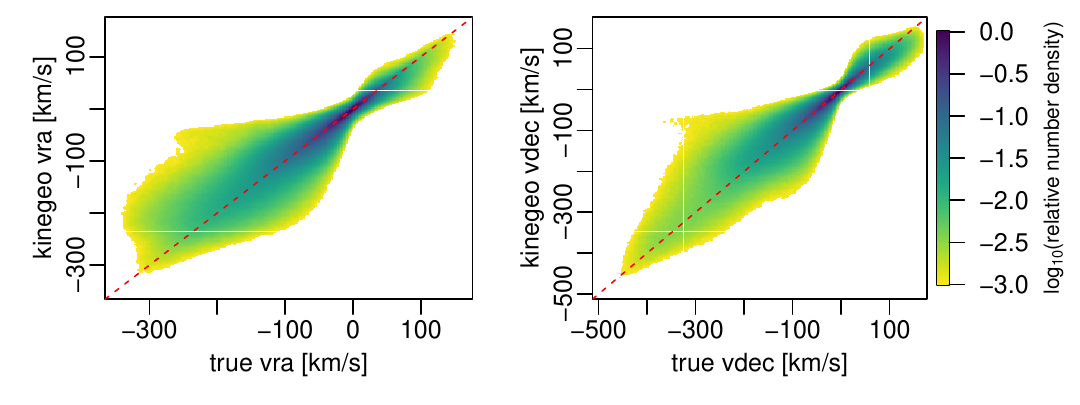}
  \caption{Estimated distances and velocities vs their true values for the constant fraction sample in the mock catalogue.
    The colour scale denotes the density of points on a logarithmic scale relative to the maximum in that panel, with densities less than a thousandth of the peak in white.
\label{fig:results_mock_est_vs_true_density}}
\end{center}
\end{figure*}

Figure~\ref{fig:results_mock_est_vs_true_density} compares the estimated median distances and velocities against their true values for the constant fraction sample. The top left panel shows the geometric distance estimated from equation~\ref{eqn:distpost_geo}.  The other panels show the kinegeometric distance and two velocity components (\vra, \vdec) from equation~\ref{eqn:kinegeopost}. The variations and differences are discussed in more detail below. Notable for now, though, is the smaller spread for kinegeometric distances compared to geometric ones at larger distances.
From now on I will discuss and show results for the total kinegeometric velocity $\sqrt{\vra^2 + \vdec^2}$ rather than its individual components. This will be compared with the total velocity derived from the geometric method, $kr_{\rm geo}\sqrt{\pmra^2 + \pmdec^2}$.

\begin{table*}
\begin{center}
\caption{Median performance for the two distance and two velocity
  estimates according to four different statistics, computed on the \gdr3\ mock catalogue using the two
  different samples (section~\ref{sec:results_mock}) in the top and middle blocks.  The bottom block shows the results for simulations at the higher precisions expected in \gdr5, discussed in section~\ref{sec:gdr5}.
\label{tab:results_mock_resid_statistics}
}
\begin{tabular}{rrrrrr}
  \hline
                                  & & geo distance & kinegeo distance & geo velocity & kinegeo velocity \\
  \hline
                                 \multicolumn{2}{r}{{\em constant fraction sample}} & & & & \\

bias   & est-true                                 & $-20.9$\,\pc  & $-29.9$\,\pc & $-0.410$\,\kms   &  $-0.875$\,\kms  \\
scatter & $\mid$ est-true $\mid$              & $977$\,\pc  & $779$\,\pc  & $19.8$\,\kms  & $16.0$\,\kms   \\
fractional bias &   (est-true)/true                          & $-0.0128$   & $-0.0156$  &  $-0.0098$   &  $-0.0178$   \\
fractional scatter & $\mid$ (est-true)/true $\mid$     & 0.228    &  0.185   &  0.238   & 0.192       \\
  \hline
                                 \multicolumn{2}{r}{{\em constant number sample}} & & & & \\
bias   &    est-true                                 & $-0.788$\,\pc     &  $-2.53$\,\pc    &  $-0.030$\,\kms &  $-0.192$\,\kms  \\
scatter & $\mid$ est-true $\mid$        & $199$\,\pc   &  $177$\,\pc  &   $6.14$\,\kms   &   $5.68$\,\kms \\    
fractional bias &  (est-true)/true          & $-0.0018$    & $-0.0047$      &  $-0.0009$ &   $-0.0055$        \\
fractional scatter &  $\mid$ (est-true)/true $\mid$  &  0.125    &   0.112      &   0.128  &   0.114       \\
  \hline
  \hline
                                  \multicolumn{2}{r}{{\em constant fraction sample}} &  \multicolumn{4}{l}{for expected \gdr5\ astrometric precisions} \\

bias   & est-true                                          & $-10.1$\,\pc          & $-18.1$\,\pc & $-0.229$\,\kms   &  $-0.461$\,\kms  \\
scatter & $\mid$ est-true $\mid$             & $722$\,\pc              & $595$\,\pc  & $14.7$\,\kms  & $11.6$\,\kms   \\
fractional bias &   (est-true)/true                               & $-0.0072$   & $-0.0105$  &  $-0.0057$   &  $-0.0106$   \\
fractional scatter & $\mid$ (est-true)/true $\mid$     & 0.168        &  0.139        &  0.178            & 0.139       \\
  \hline
\end{tabular}
\end{center}
\end{table*}

Performance statistics for the constant fraction sample are summarized in the top block of Table~\ref{tab:results_mock_resid_statistics}.
For the distances I use the fractional residuals, (estimated-true)/true, because distance accuracy is a strong function of distance.
For the total velocities I use the actual residuals, (estimated-true).
We see that the residuals for the two methods are broadly similar, with a slightly heavier tail to negative residuals.
For the distances, the fractional {\em bias} -- median of (estimated-true)/true -- is $-0.0128$ for the geometric distance and $-0.0156$ for the kinegeometric distance.
For the velocities, the bias -- median of (estimated-true) -- is $-0.41$\,\kms\ for the geometric method and $-0.88$\,\kms\ for the kinegeometric one.
Thus the biases are quite small on average.

The biases, which we can think of as the systematic deviation about
the identity line in Figure~\ref{fig:results_mock_est_vs_true_density}, are part of the estimation error.
The total estimation error I characterize by the median of the absolute values of the residuals, more specifically 
as median($|$estimated-true)$|$/true) for the distance and median($|$estimated-true$|$) for the velocities.
These robust versions of the error I refer to as the {\em scatter}. Note that it includes the bias.
For the distances, the fractional scatters are $0.228$ for geometric distances and $0.185$ for kinegeometric distances.
For the velocities, the scatters are $19.8$\,\kms\ for the geometric method and $16.0$\,\kms\ for the kinegeometric one.
These are quite large, but it must be remembered that most of the stars in the sample are distant and faint.

The middle block of Table~\ref{tab:results_mock_resid_statistics} shows the performance statistics on the constant number sample, where we see much smaller bias and scatter in all cases.
This is because this sample has more stars at higher latitudes that are generally nearer.
This is not as representative, so for global statistics I report the worse results on the constant number sample elsewhere in this paper. The final block is for \gdr5, discussed in section~\ref{sec:gdr5}.

The goal of the present paper is to introduce the kinegeometric estimates. The above statistics show
that kinegeometric distances are overall more accurate than the geometric ones by
a factor of only 1.25 (the ratio of the scatters).
Within this they have a bias 1.2 times larger than the geometric estimates,
although the average bias is only $-1.5$\%.
In the constant fraction sample, 58\% of the stars have a kinegeometric error (scatter) that is smaller than the geometric error. 27\% of stars have a kinegeometric error which is smaller than the geometric error by a factor of two or more (compared to 17\% the other way around). The is no obvious, narrow part of parameter space where one estimate is always better than the other, but there are broad variations with distance and position, which we now explore.

\subsection{Performance as a function of distance}

\begin{figure*}[h]
\begin{center}
  \includegraphics[width=0.49\textwidth, angle=0]{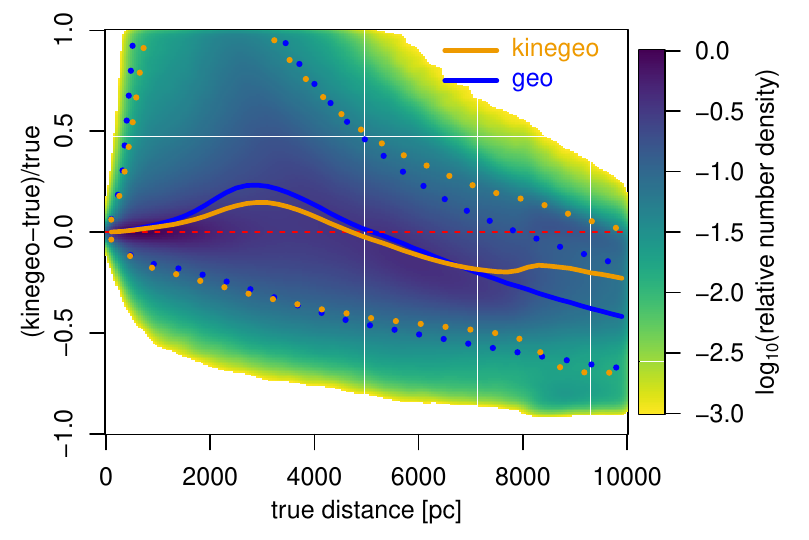}
  \includegraphics[width=0.49\textwidth, angle=0]{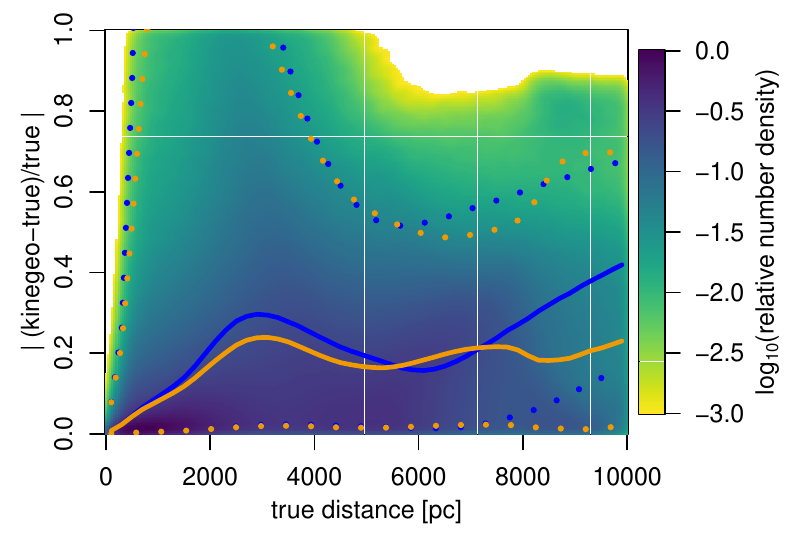}
  \caption{Performance of the kinegeometric distance estimate (median of the posterior) as a function of distance for the constant fraction sample in the mock catalogue. The left panel shows the fractional bias in the estimates, the right panel the fractional scatter. The colour scale shows the density of stars on a log scale. The solid orange line shows the median bias or scatter at each distance, the dotted lines are the 5th and 95th percentiles. The blue lines show the same metrics for the geometric distances for comparison. 
\label{fig:results_mock_rMedKinogeo_bias_scatter_vs_distance}}
\end{center}
\end{figure*}

\begin{figure*}[h]
\begin{center}
  \includegraphics[width=0.49\textwidth, angle=0]{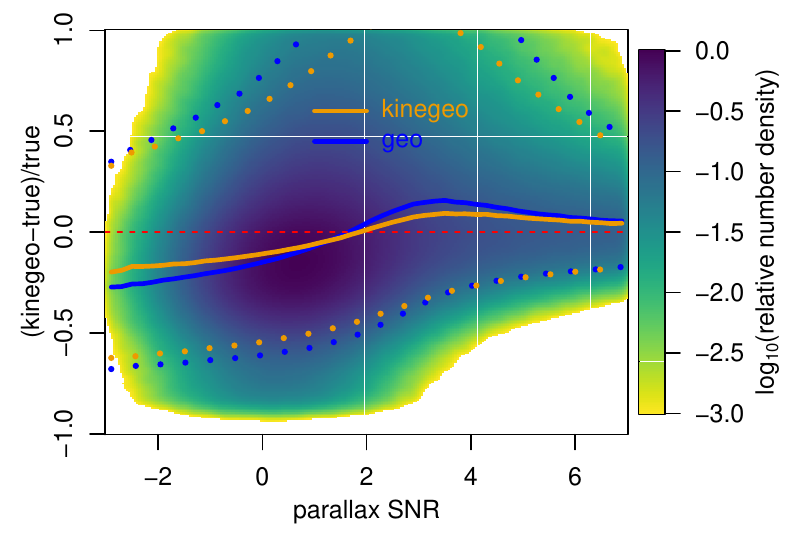}
  \includegraphics[width=0.49\textwidth, angle=0]{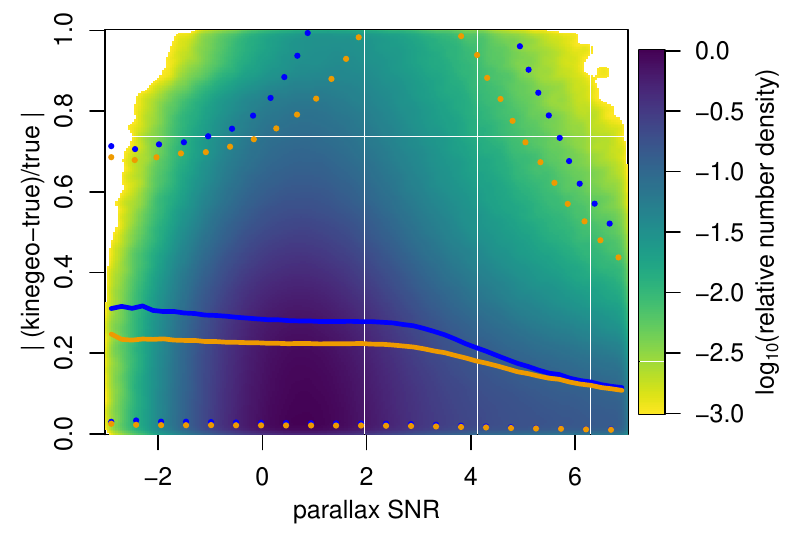}
  \caption{As Figure~\ref{fig:results_mock_rMedKinogeo_bias_scatter_vs_distance}, but now as a function of parallax SNR.
\label{fig:results_mock_rMedKinogeo_bias_scatter_vs_parallaxSNR}}
\end{center}
\end{figure*}

Figure~\ref{fig:results_mock_rMedKinogeo_bias_scatter_vs_distance} shows how the bias and scatter of the kinegeometric distance estimate varies as a function of distance as a density plot, with the orange line indicating the median at each distance.
We see a complex dependence on distance, in part because it is an average over sources over the whole sky.
Compared to the geometric method (blue line),
the kinegeometric bias and scatter are slightly smaller at most distances. The reason the geometric distance nonetheless showed a slightly smaller bias overall (Table~\ref{tab:results_mock_resid_statistics}) is because a large number of stars lie in the range where the geometric bias is smaller. Beyond about 7\,\kpc\ the kinegeometric distances are more accurate. This is the regime where the kinegeometric distances are most useful.

Figure~\ref{fig:results_mock_rMedKinogeo_bias_scatter_vs_parallaxSNR} shows the performance as a function of parallax SNR (defined as the parallax divided by the parallax uncertainty).
The performances of the two methods are similar in terms of bias, but the kinegeometric distances have lower scatter for low parallax SNR.
This is precisely the region where the proper motions are helping, because even though the proper motion SNRs are generally lower too, they are still high enough to provide additional information on distance. At higher parallax SNR the differences are small because the parallax dominates, and the priors are less important.

\begin{figure*}
\begin{center}
  \includegraphics[width=0.49\textwidth, angle=0]{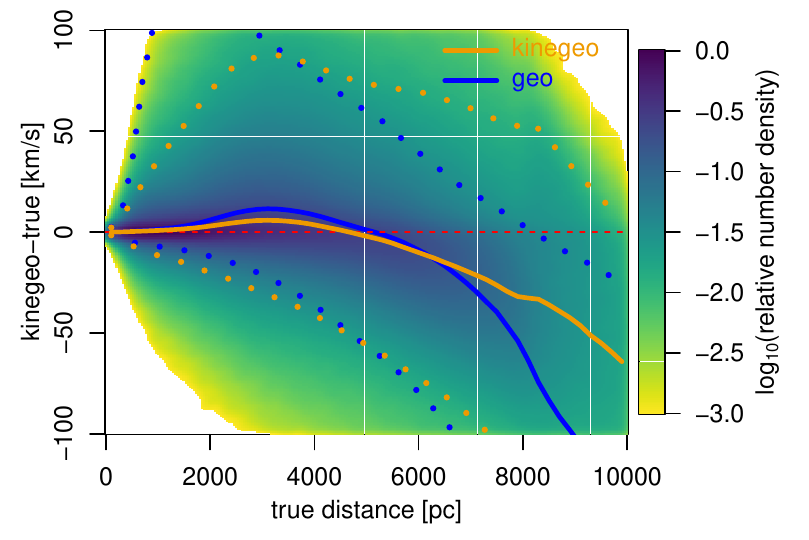}
  \includegraphics[width=0.49\textwidth, angle=0]{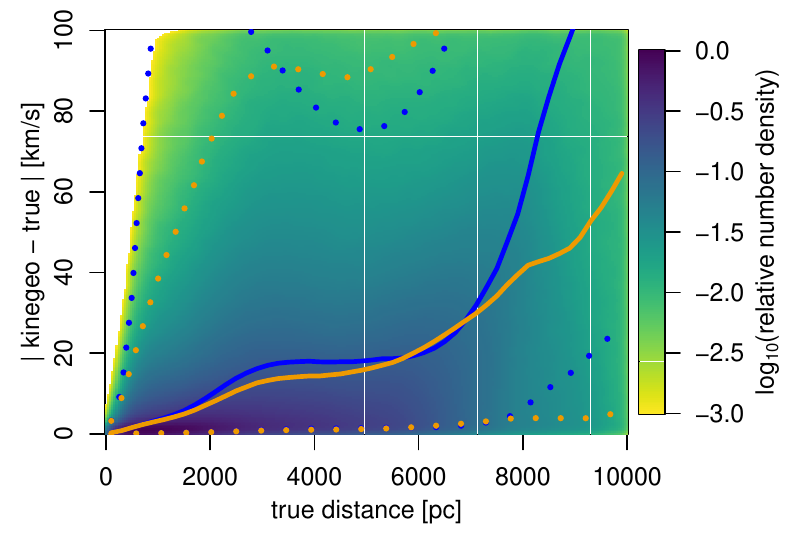}
  \caption{Performance of the kinegeometric velocity estimate (median of the posterior) as a function of distance for the constant fraction sample in the mock catalogue. The left panel shows the bias in the estimates, the right panel the scatter. The colour scale shows the density of stars on a log scale. The solid orange line shows the median bias or scatter at each distance, the dotted lines the 5th and 95th percentiles. The blue lines show the same metrics for the geometric method (geometric distance times proper motion).
\label{fig:results_mock_vtotMedKinogeo_bias_scatter_vs_distance}}
\end{center}
\end{figure*}

Figure~\ref{fig:results_mock_vtotMedKinogeo_bias_scatter_vs_distance} compares the bias and scatter of the two total velocity estimates as a function of distance. Again we see a large range of these statistics at a given distance, as evidence by the dotted lines in these plots. The velocity estimates have similar performances (in both bias and scatter) for stars with true distances less than about 7\,\kpc. At larger distance the kinegeometric velocity estimates are more accurate, due to the assistance from the proper motions and velocity prior.


\subsection{Performance as a function of direction}

We turn look at how the performance varies over the sky using the constant number sample.

\begin{figure*}
\begin{center}
  \includegraphics[width=0.49\textwidth, angle=0]{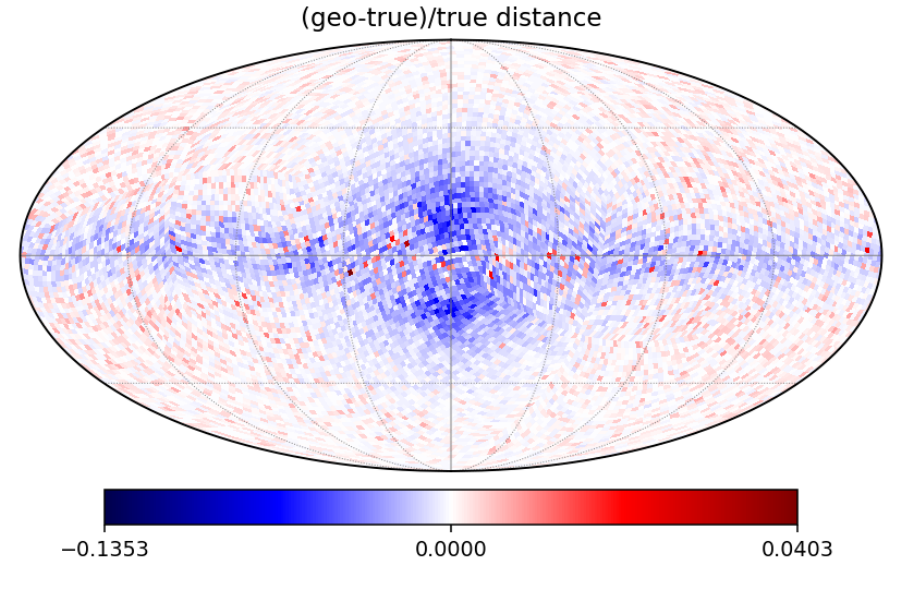}
  \includegraphics[width=0.49\textwidth, angle=0]{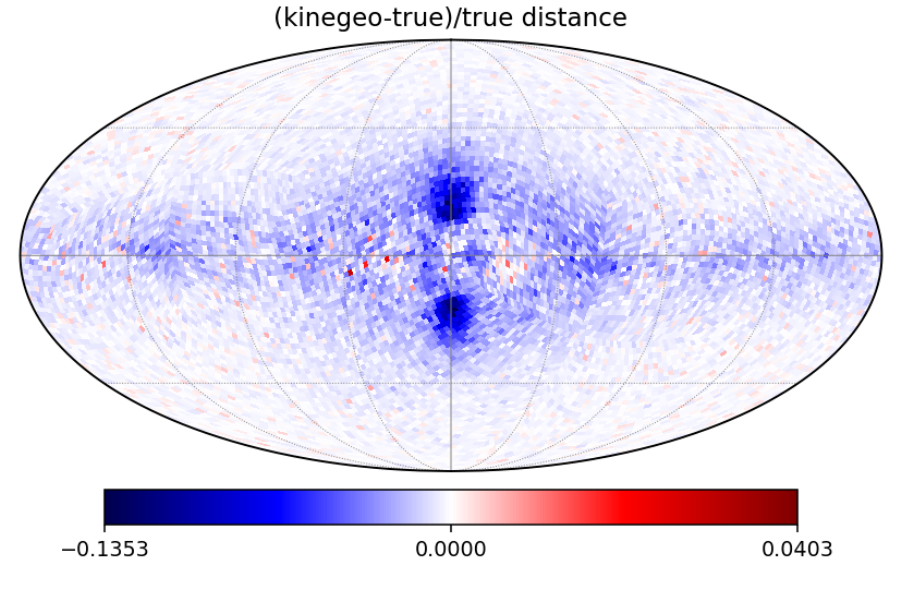}
  \includegraphics[width=0.49\textwidth, angle=0]{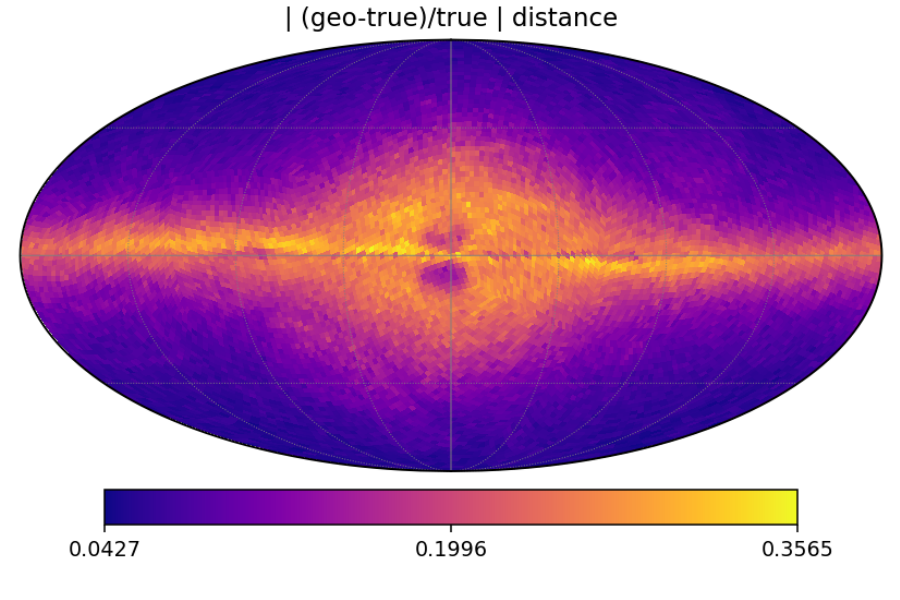}
  \includegraphics[width=0.49\textwidth, angle=0]{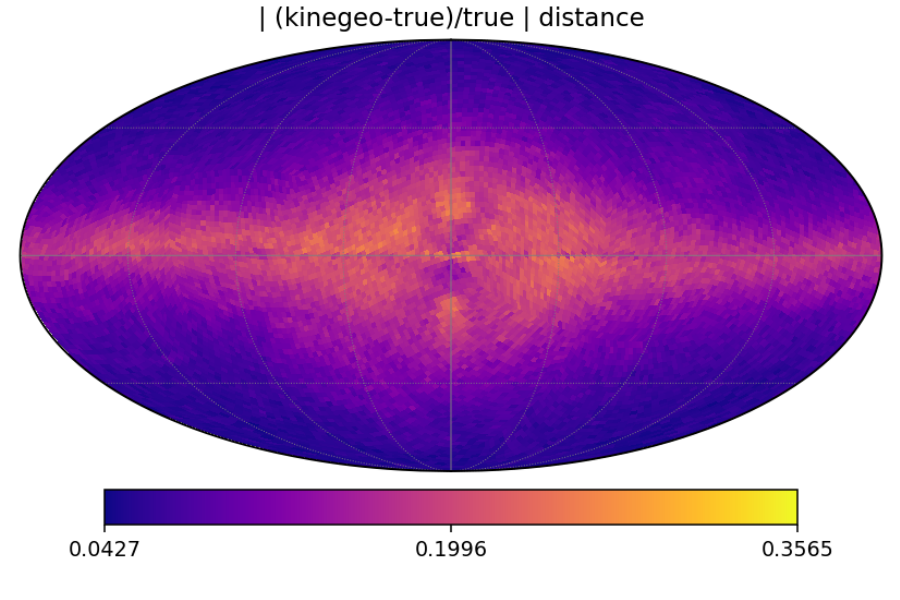}
  \caption{Fractional distance residuals -- (estimated-true)/true -- per HEALpixel for the constant number sample in the mock catalogue. The top row shows the median fractional residual (the bias); the bottom row shows the median absolute fractional residual (the scatter). The left column is for geometric estimates, the right column is for kinegeometric estimates.
    The colour bars span the full common range in each of the top and bottom rows.
   The top row uses a bilinear colour bar (separate scales for negative and positive values): The dominance of red HEALpixels over blue ones at high latitudes is because the positive red scale extends to smaller values than the negative blue scale; the absolute biases are more or less the same overall.
    \label{fig:results_mock_rMedGeo_rMedKinogeo_bias_scatter_skyplot}
  }
\end{center}
\end{figure*}

Figure~\ref{fig:results_mock_rMedGeo_rMedKinogeo_bias_scatter_skyplot} shows the fractional bias and scatter, by computing the statistic for each star and then taking the median over each HEALpixel.
Looking first at the upper panels (the bias), we see that both methods tend to underestimate distances in lines-of-sight in the disk and bulge. This is because these directions are dominated by distant stars (see Figure~\ref{fig:distance_prior_median_skyplot}) which generally have low parallax SNRs, which in turn introduces a bias (explained in section~\ref{sec:cause_of_bias}).
The kinegeometric estimates show larger negative biases by up to an average of 13\% in two regions above and below the Galactic centre. This is related to the large changes in the velocity prior beyond a few \kpc\ (see Figure~\ref{fig:veltotmeanprior_skyplot}). The geometric distance estimates are less affected.

The scatter plots (bottom row of Figure~\ref{fig:results_mock_rMedGeo_rMedKinogeo_bias_scatter_skyplot}) show a larger scatter in both distance estimates at low latitudes. This is a consequence of most stars being at large true distances, so have low SNR parallaxes. The kinegeometric distances show a smaller scatter though, because
the proper motions improve the distance accuracy primarily at low parallax SNR (as already discussed in the context of Figure~\ref{fig:results_mock_rMedKinogeo_bias_scatter_vs_parallaxSNR}).
 
\begin{figure}
\begin{center}
  \includegraphics[width=0.49\textwidth, angle=0]{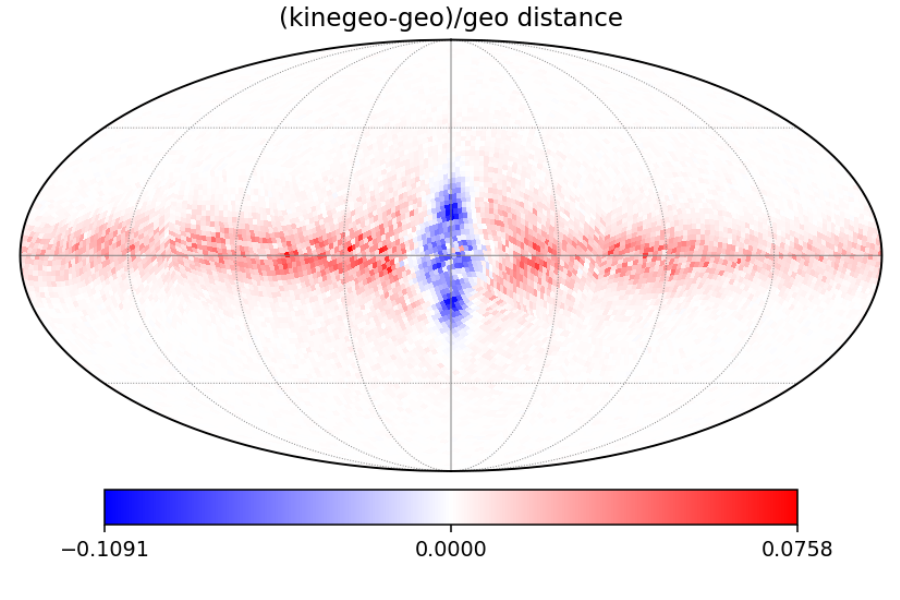}
  \caption{Median fractional differences between the kinegeometric and geometric distances per HEALpixel (computed on the individual stars),
    for the constant number sample in the mock catalogue.
    The colour bar has white at zero and diverges on different linear scales to the maximum negative and positive differences. 
\label{fig:results_mock_cns_rMedGeo_rMedKinogeo_fracdiff_skyplot}}
\end{center}
\end{figure}

Figure~\ref{fig:results_mock_cns_rMedGeo_rMedKinogeo_fracdiff_skyplot} shows the differences between the kinegeometric and geometric distance estimates over the sky
(computing the median per HEALpixel of the differences for each star).
At higher Galactic latitudes the estimates are very similar on average. In the central regions of the Galaxy the kinegeometric distances are smaller, by up to 11\% on average within a HEALpixel. In the rest of the Galactic disk the kinegeometric distances are larger by up to 8\% on average.

\begin{figure*}
\begin{center}
  \includegraphics[width=0.49\textwidth, angle=0]{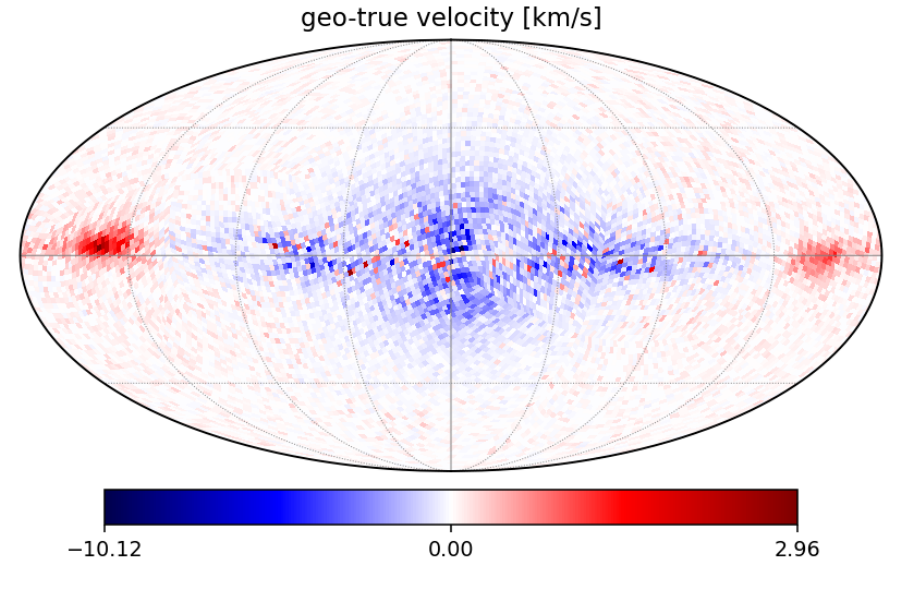}
  \includegraphics[width=0.49\textwidth, angle=0]{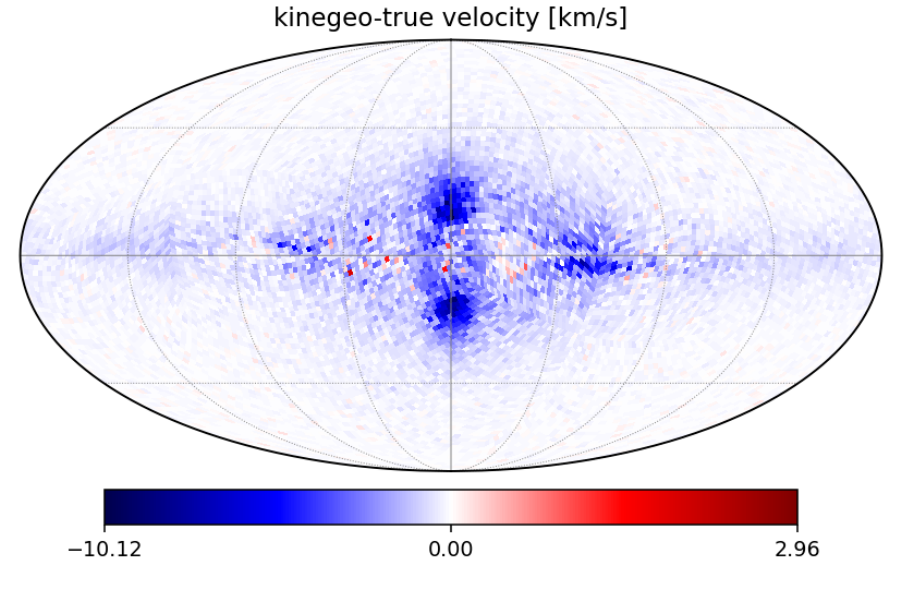}
  \includegraphics[width=0.49\textwidth, angle=0]{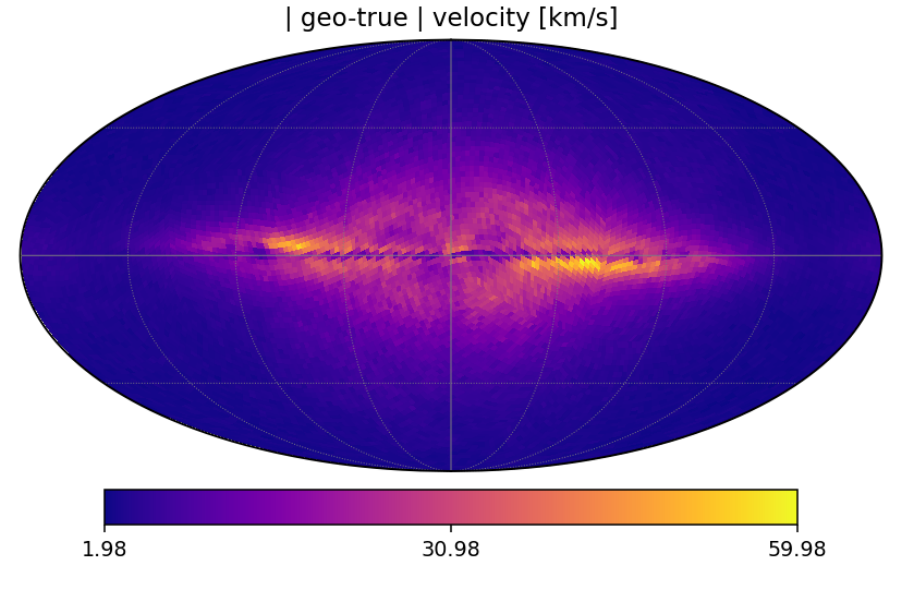}
  \includegraphics[width=0.49\textwidth, angle=0]{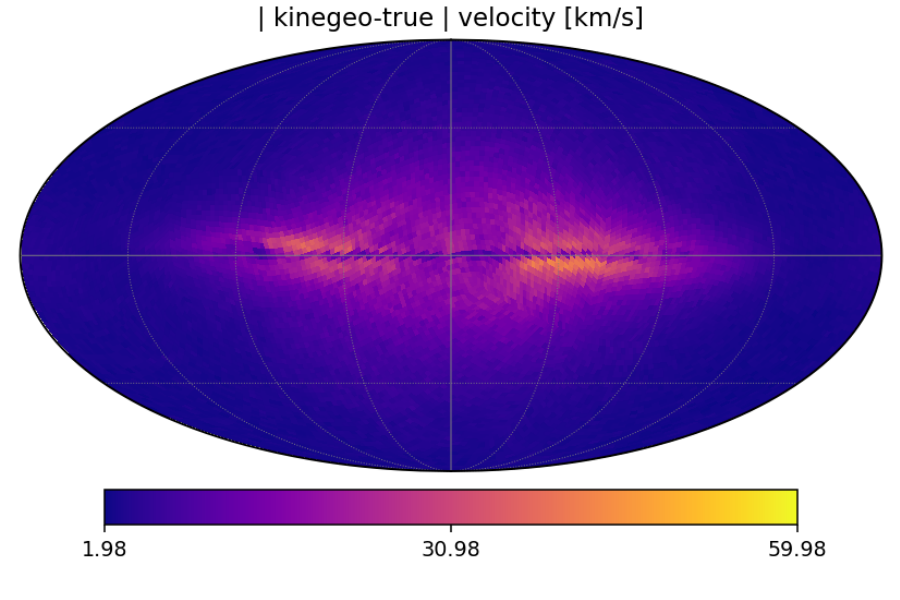}
  \caption{Velocity residuals -- (estimated-true) --  per HEALpixel for the constant number sample in the mock catalogue. The top row shows the median residual (the bias); the bottom row shows the median absolute residual (the scatter). The left column is for geometric method (geometric distance times proper motion), the right column is for kinegeometric estimates.
   The colour bars span the full common range in each of the top and bottom rows. 
   The top row uses a bilinear colour bar (separate scales for negative and positive values).
\label{fig:results_mock_vtotDirect_vtotMedKinogeo_bias_scatter_skyplot}}
\end{center}
\end{figure*}

Figure~\ref{fig:results_mock_vtotDirect_vtotMedKinogeo_bias_scatter_skyplot} shows how the performances of the two velocity estimates vary over the sky. The bias patterns are broadly similar to what we saw with the distances, especially for the kinegeometric estimates, e.g.\ the blue blobs at $\glon \simeq 0\degree$ mentioned above.
The bias map for the geometric velocities, however, shows two regions in the disk with a small but significant overestimate, at around $\glon = -155\degree$ and $\glon = 150\degree$ (the ``red cheeks'' in the top-left panel of Figure~\ref{fig:results_mock_vtotDirect_vtotMedKinogeo_bias_scatter_skyplot}). They are slightly displaced from the Galactic plane, so are presumably related to the warp of the disk in the Galaxy model on which the mock catalogue is based. They do not appear in the bias map for kinegeometric distances (top-right panel of Figure~\ref{fig:results_mock_vtotDirect_vtotMedKinogeo_bias_scatter_skyplot}). Some structures in these regions are seen in the nearer slices of the velocity prior  (Figure~\ref{fig:veltotmeanprior_skyplot}), but not in the distance prior (Figure~\ref{fig:distance_prior_median_skyplot}), suggesting that
not accounting for these in the geometric prior is problematic.
Naturally, if these features do not exist in the real Galaxy, the (unknown) biases in the inference in \gdr3\ could be different.

\begin{figure}
\begin{center}
  \includegraphics[width=0.49\textwidth, angle=0]{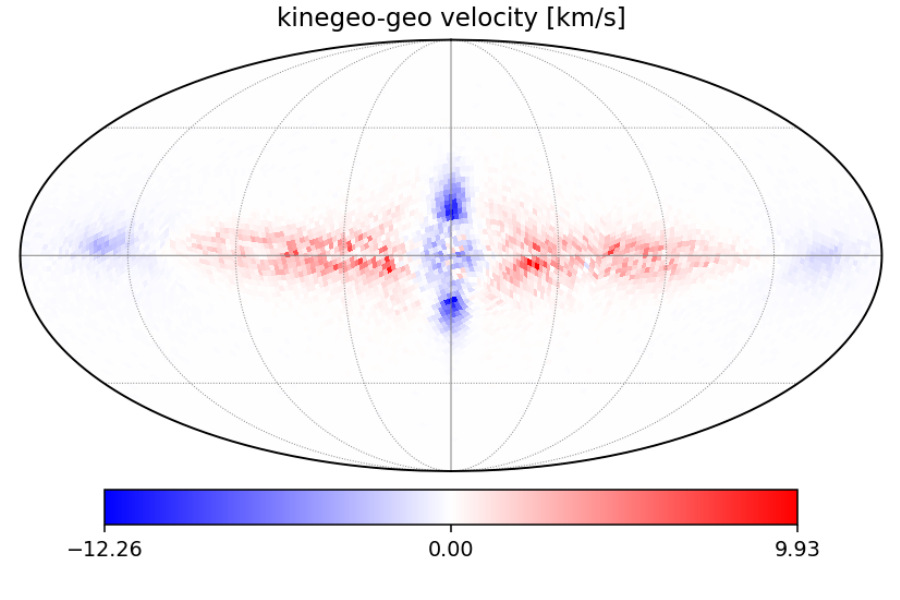}
  \caption{Median differences between the kinegeometric and geometric velocities per HEALpixel (computed on the individual stars),
  for the constant number sample in the mock catalogue. The colour bar has white at zero and diverges on different linear scales
  to the maximum negative and positive differences. 
\label{fig:results_mock_cns_vtotDirect_vtotKinogeo_diff_skyplot}}
\end{center}
\end{figure}

The lower panels of Figure~\ref{fig:results_mock_vtotDirect_vtotMedKinogeo_bias_scatter_skyplot} show the scatter in the two velocity estimates, which are similar in pattern, but slightly larger for geometric velocities.
Figure~\ref{fig:results_mock_cns_vtotDirect_vtotKinogeo_diff_skyplot} shows the average differences per HEALpixel between the two velocity estimates.

\subsection{Uncertainty estimates}\label{sec:results_mock_uncertainty_estimates}


\begin{figure}
\begin{center}
  \includegraphics[width=0.49\textwidth, angle=0]{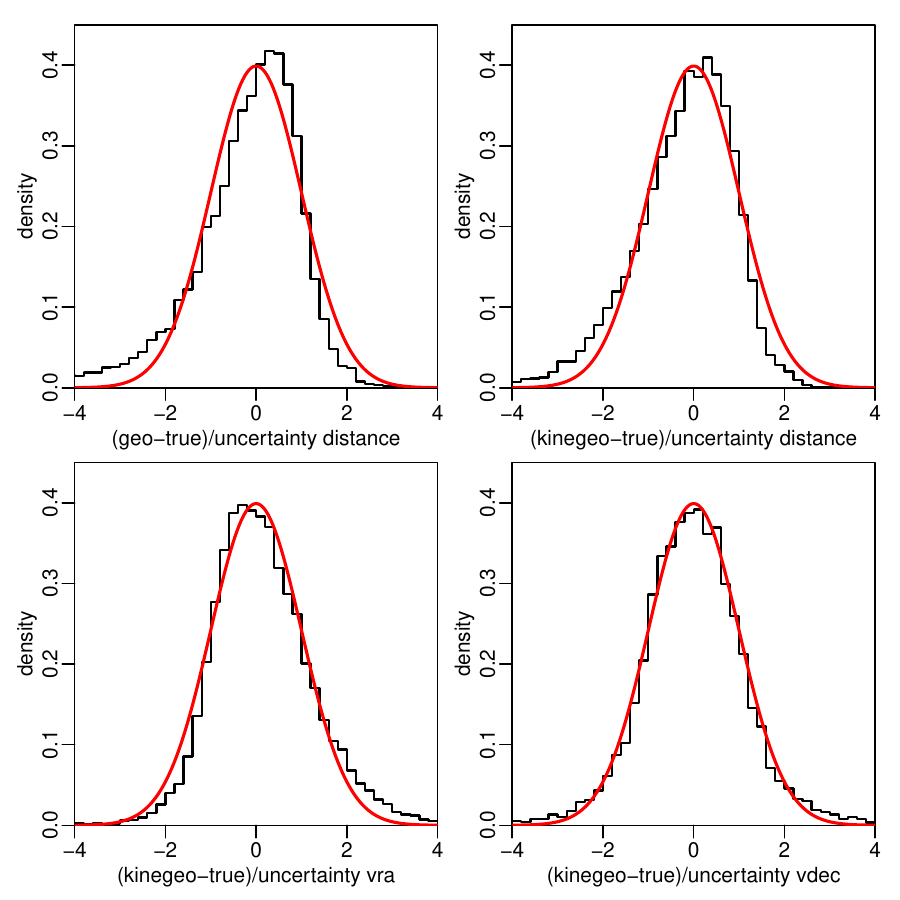}
\caption{Normalized residuals for the geometric posterior distance (top left) and for the three parameters of the kinegeometric posterior (other panels), computed for all sources in mock HEALpixel 6200. The smooth red lines show the unit Gaussian for comparison. \label{fig:results_mock_06200_normalized_residual_distributions}}
\end{center}
\end{figure}

From the posterior PDF for each star I compute the 16th and 84th percentiles for each parameter, then take their difference from the median to obtain lower and upper 1$\sigma$-like uncertainty estimates. Half their difference gives a single uncertainty estimate for the parameter.
If the residuals (median minus true) have an unbiased Gaussian distribution, and if the uncertainty is a statistically correct estimate of the residual, then the ratio of the residual to the uncertainty -- the normalized residuals -- should have a unit Gaussian distribution.
These are shown in Figure~\ref{fig:results_mock_06200_normalized_residual_distributions}. All four distributions have a mean close to zero and a standard deviation close to unity, although the distributions for both distance measures show a slight negative skew. This is not surprising because we know there is a negative distance bias for distant, low parallax SNR stars (see appendix~\ref{sec:cause_of_bias}). But overall this shows that the uncertainty estimates are good statistical estimates of the true error.

\vspace*{1em}
\section{Results on \release}\label{sec:results}


1.46 bilion stars in \release\ have parallaxes and proper motions.
As with the mock catalogue, I apply the geometric and kinegeometric inference methods to two randomly-selected subsets:
a constant number sample
with 0.8\% of all sources (10.3 million sources),
and a constant fraction sample with 900 sources per HEALpixel (10.8 million sources),
in both cases restricted to sources brighter than the prior magnitude limit in each HEALpixel
\citep[][section 1]{2021AJ....161..147B} to enable comparison to the estimates on the mock catalogue in the previous section.

\subsection{Distances}

\begin{figure}
\begin{center}
\includegraphics[width=0.49\textwidth, angle=0]{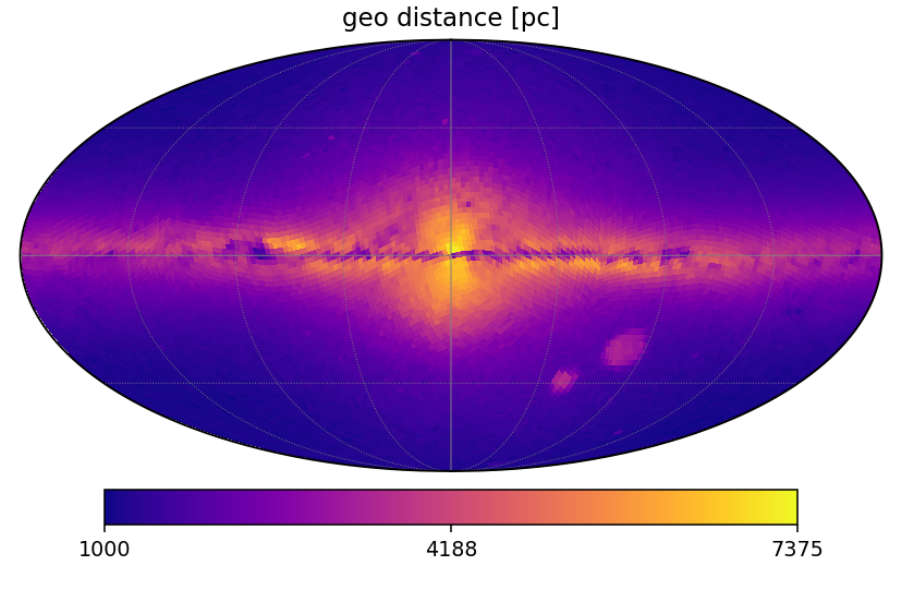}
\includegraphics[width=0.49\textwidth, angle=0]{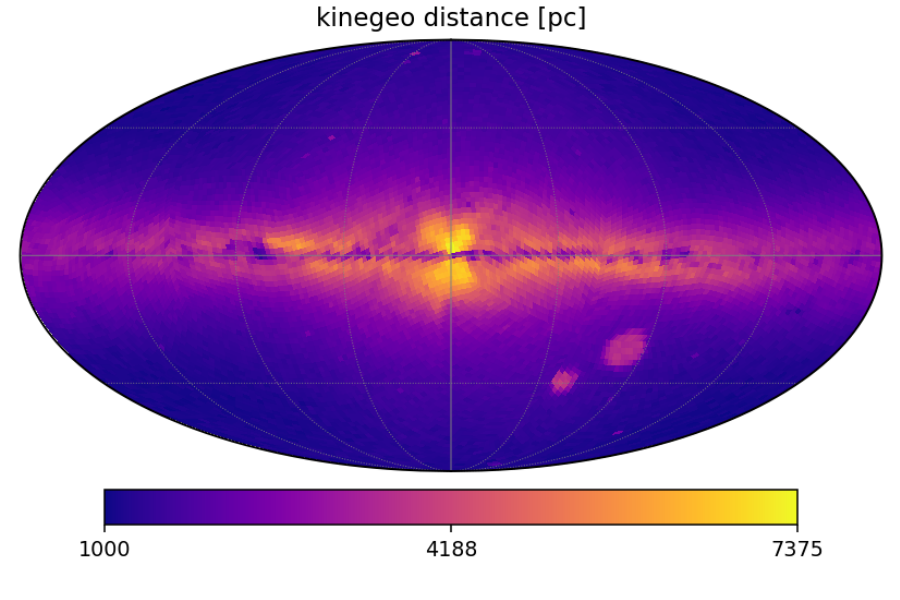}
\includegraphics[width=0.49\textwidth, angle=0]{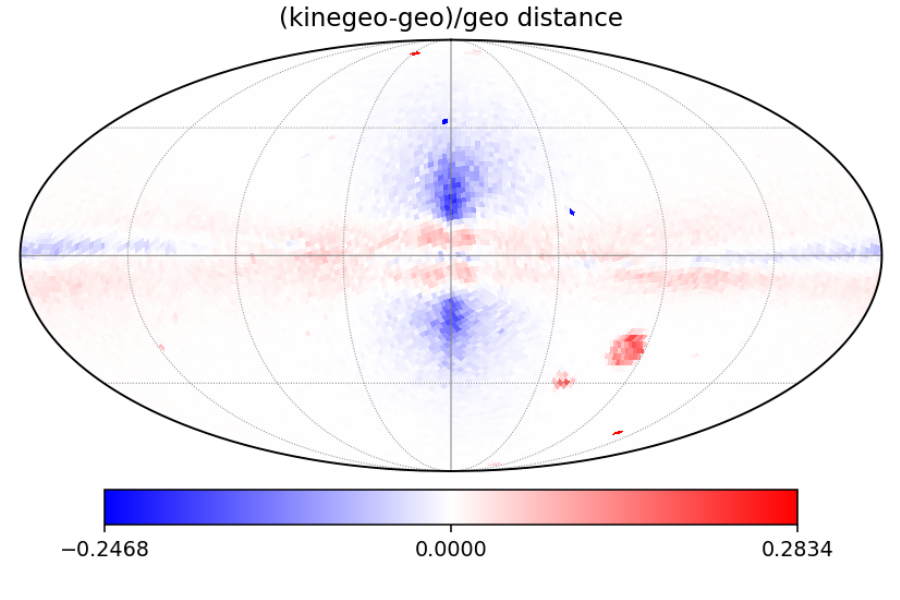}
\caption{Median distance per HEALpixel for the constant number sample in \release: Geometric (top) and kinegeometric (middle).
  The linear colour bar spans the full common range. The bottom panel shows
  the median fractional differences between these (computed on the individual stars). It
uses a bilinear colour bar with white at zero and diverges on different linear scales to the maximum negative and positive differences.
\label{fig:results_rMedGeo_rMedKinogeo_skyplot}}
\end{center}
\end{figure}

Figure~\ref{fig:results_rMedGeo_rMedKinogeo_skyplot} shows the median distance per HEALpixel.
The median distance increases to lower Galactic latitudes in both methods, because of the higher density of stars at larger distances within the disk. The median distance is smaller
in the Galactic plane, however, because \gaia's depth is limited by strong interstellar dust extinction. Many of the patterns we see close to the plane are due to dust density variations. At high Galactic latitudes the depth through the disk is small, and although distant stars in the halo are visible, they are rare, so the median remains relatively small.
The bulge is quite prominent in both distance estimates as a larger, approximately circular region of more distant stars. Close to the Galactic centre we see dust-free lines-of-sight that extend the median distance to over 7\,kpc.
Neither the mock catalogue as we use it nor the prior contains star clusters, so cluster distances will not be correct. This is most noticeable for stars in the Large and Small Magellanic Clouds, which we see have severely underestimated distances.

The differences between the two distance estimates are shown in the bottom panel of Figure~\ref{fig:results_rMedGeo_rMedKinogeo_skyplot} (median of differences per star). The largest differences are in parts of the Galactic plane, parts of the bulge, and in several star clusters.
Most prominent is that the kinegeometric distance is smaller in the bulge above and below the Galactic centre. This is a result of the measured proper motions combined with the velocities of the Galaxy prior.
In parts of the Galactic plane the kinegeometric distances are larger, in others the differences are negligible, and in the northern part of the disk near the anticentre the kinegeometric distances are smaller. These patterns are presumably related to the Galactic bar and warp, which unsurprisingly differ between \gdr3\ and the Galactic model used to make the prior.

The larger kinegeometric distances in the upper and lower parts of the bulge was also seen in the results on the noisy mock data in Figure~\ref{fig:results_mock_cns_rMedGeo_rMedKinogeo_fracdiff_skyplot}, although it was less prominent there and also extended through the Galactic disk.
  We should not over interpret differences between Figures~\ref{fig:results_mock_cns_rMedGeo_rMedKinogeo_fracdiff_skyplot} and~\ref{fig:results_rMedGeo_rMedKinogeo_skyplot}, however: they can occur for many reasons, including differences in the underlying stellar spatial and velocity distributions or in the extinctions between the mock catalogue and real Galaxy.
The larger colour bar scale on the difference plot for the \gdr3\ results shows that kinegeometric and geometric distances differ much more in the real inference than they did in the mock inference, no doubt a result of these differences.

\begin{figure}
\begin{center}
\includegraphics[width=0.49\textwidth, angle=0]{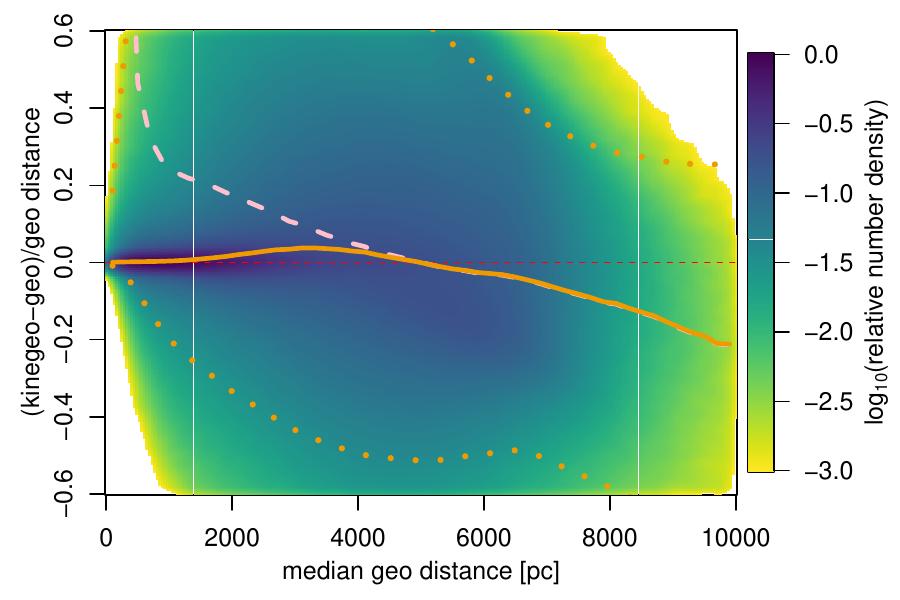}
\caption{Variation of the fractional difference between the kinegeometric and geometric distances with geometric distance for the constant fraction sample in \gdr3.
The colour scale shows the density of stars on a log scale. The solid orange line shows the median fractional difference, the dotted lines the 5th and 95th percentiles.  The dashed pink line shows the median fractional difference for just the subset with parallax SNR less than three.
\label{fig:results_cfs_kinogeo_minus_geo_over_geo_vs_geo}}
\end{center}
\end{figure}

Figure~\ref{fig:results_cfs_kinogeo_minus_geo_over_geo_vs_geo} compares the two distance estimates as a function of geometric distance. The density plot and the quantiles (dashed lines) show that there is a large spread, but the median trend (solid orange line) is for the
kinegeometric distances to be on average smaller than the geometric ones beyond 6\,\kpc, by 20\% at 10\,\kpc.\footnote{One might read the left panel of Figure~\ref{fig:results_mock_rMedKinogeo_bias_scatter_vs_distance} to suggest we had the opposite in the mock data, but that plot shows the biases against the {\em true} distance.
If we replace true with geometric, in then looks very similar to Figure~\ref{fig:results_cfs_kinogeo_minus_geo_over_geo_vs_geo}.}
This is not true in all lines-of-sight, however, such as the Galactic centre and parts of the disk, as we saw in Figure~\ref{fig:results_rMedGeo_rMedKinogeo_skyplot}.
When we consider just the low parallax SNR stars ($< 3$), we see a large difference in the distance estimates on average also for nearby stars (the dashed pink line in Figure~\ref{fig:results_cfs_kinogeo_minus_geo_over_geo_vs_geo}).

\begin{figure}
\begin{center}
\includegraphics[width=0.49\textwidth, angle=0]{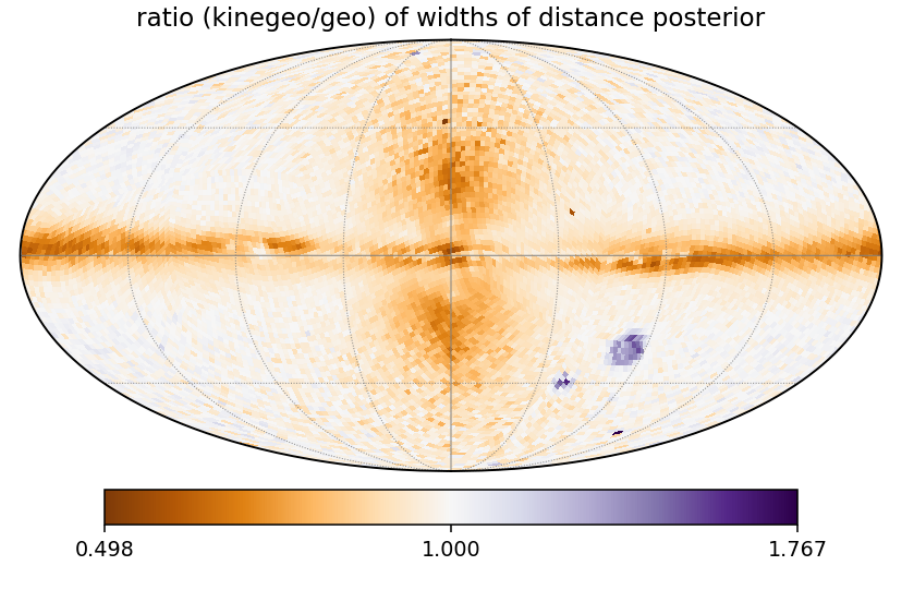}
\caption{Median per HEALpixel of the ratio (kinegeo/geo) of the posterior widths for the constant fraction sample in \release.
Numbers less than 1 (in orange) show where the width of the kinegeometric distance posterior is narrower than the width of the geometric distance posterior, and are therefore statistically likely to be more accurate. We see that this is the case in most parts of the sky, but particularly in the disk and bulge.
The bilinear colour bar with white at unity and diverges on different linear scales to the minimum and maximum values.
\label{fig:results_cfs_distances_pwratio_skyplot}}
\end{center}
\end{figure}

We saw in section~\ref{sec:results_mock_uncertainty_estimates} (Figure~\ref{fig:results_mock_06200_normalized_residual_distributions}) that the width of the posterior (the precision)
was a good statistical estimate of the actual error.
We can therefore compare the posterior width (the difference between the 84th and 16th percentiles) for the two types of distance estimate to predict which is more accurate. Using the constant fraction sample, we find that for 62\% of the stars, the kinegeometric posterior is narrower than the geometric one. 15\% of stars have a kinegeometric posterior that is less than half the width of the geometric one, compared to just 4\% the other way around. This suggests kinegeometric distances will be more accurate for a significant number of stars.
Figure~\ref{fig:results_cfs_distances_pwratio_skyplot} plots the ratio of these posterior widths over the sky, taking the median over all stars in each HEALpixel. We see that in almost all parts of the sky -- but particularly in the bulge and disk -- the kinegeometric distances are more precise (but remembering that each HEALpixel is an average over a large range of distances). We also find that kinegeometric distances tend to be more precise for lower parallax SNRs, as we would expect.

\subsection{Velocities}

\begin{figure}
\begin{center}
\includegraphics[width=0.49\textwidth, angle=0]{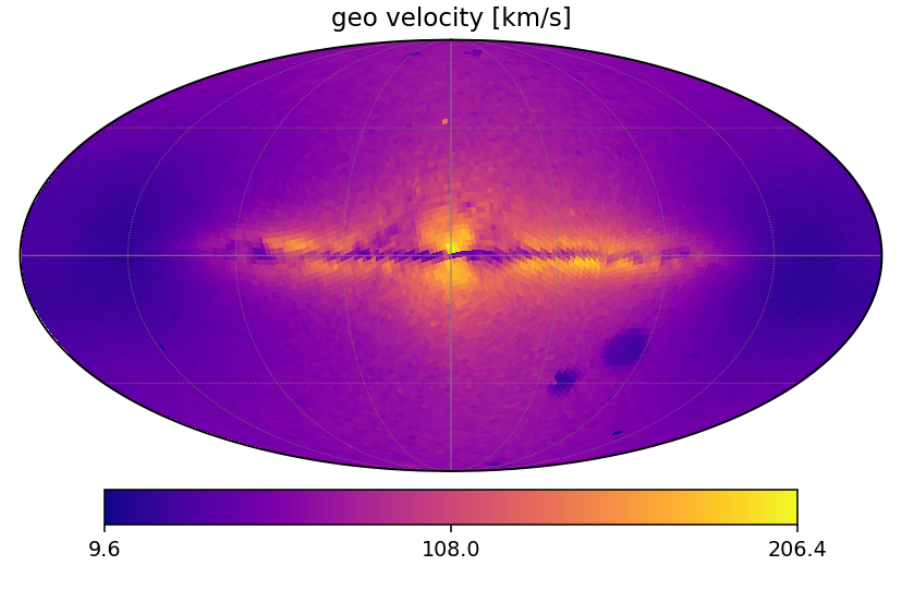}
\includegraphics[width=0.49\textwidth, angle=0]{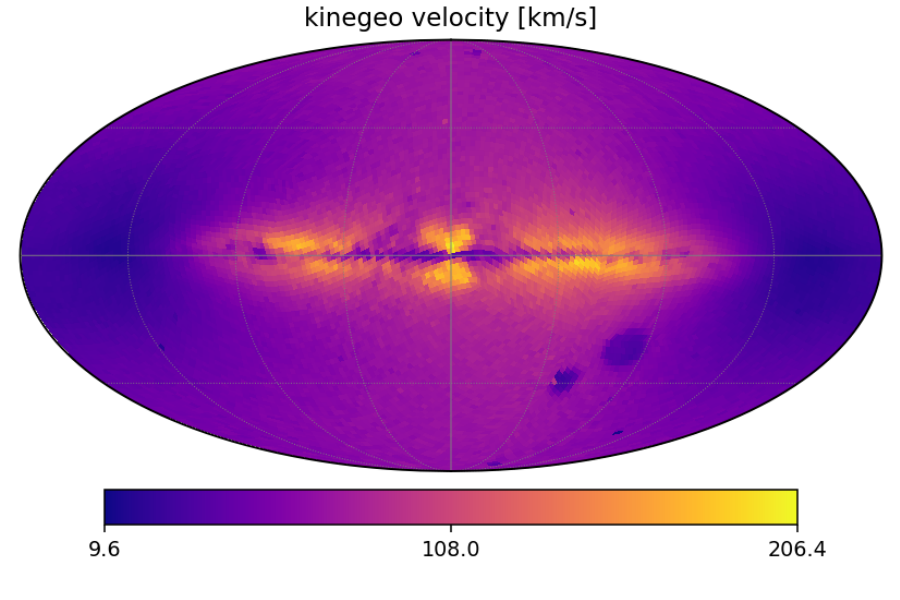}
\includegraphics[width=0.49\textwidth, angle=0]{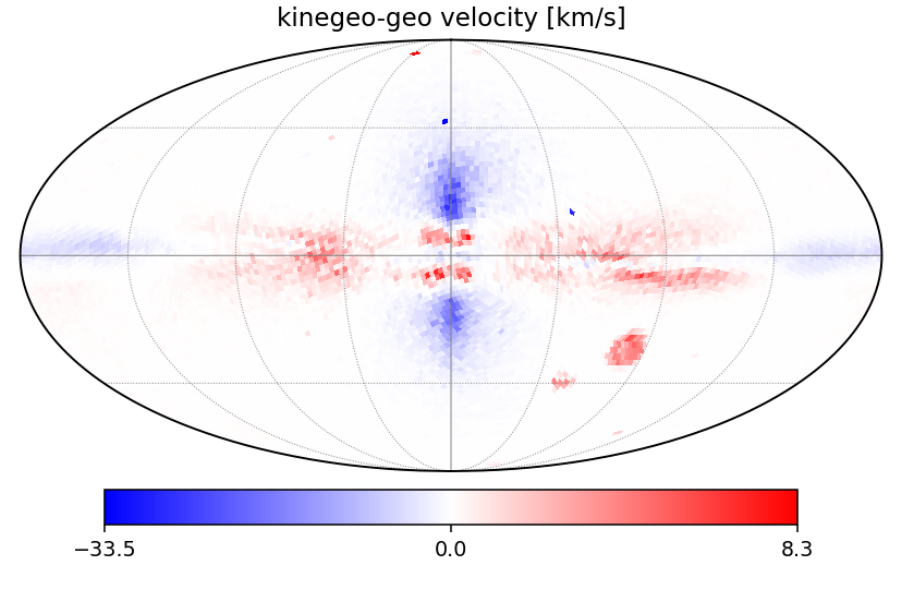}
\caption{Median transverse velocity  per HEALpixel for the constant number sample in \release: Geometric (top) and Kinegeometric (middle).
  The linear colour bar spans the full common range. The bottom panel shows
  the median differences between these (computed on the individual stars).
   It uses a bilinear colour bar with white at zero and diverges on different linear scales to the maximum negative and positive differences
\label{fig:results_vtotDirect_vtotKinogeo_skyplot}}
\end{center}
\end{figure}

Figure~\ref{fig:results_vtotDirect_vtotKinogeo_skyplot} shows the median velocity per HEALpixel inferred by both the geometric method (geometric distance times proper motion) and the kinegeometric method. These are transverse velocities relative to the solar system barycentre. The largest velocities occur within about 90\degree\ in longitude of the Galactic centre, particularly at low latitudes. This is where we see mostly disk stars on smaller radius orbits than the Sun's.
The exception is in the Galactic plane, where dust extinction means \gaia\ only sees nearby stars that are orbiting the Galaxy with low velocities relative to the Sun. 
At larger longitudes we see the outer part of the disk where, due to the flattening of the Galactic rotation curve, stars are orbiting at speeds more similar to the Sun's.

The differences between the two velocity estimates are shown in the bottom panel of Figure~\ref{fig:results_vtotDirect_vtotKinogeo_skyplot}. This is broadly similar to the difference plot obtained from the mock catalogue results in Figure~\ref{fig:results_mock_cns_vtotDirect_vtotKinogeo_diff_skyplot}, although the scales are different. The main difference (other than the absence of star clusters in the mock results), is around the Galactic centre: whereas for mock the geometric velocities were several \kms\ larger on average, in \gdr3\ the kinegeometric velocities are slightly larger.

\vspace*{1em}
\section{Expected performance with the final Gaia data release (\gdr5) and the limits of proper motions}\label{sec:gdr5}

The distance accuracy that this method can attain is determined by the accuracies of the parallaxes and proper motions, as well as the width of the distance and velocity priors.
The accuracy of the data will improve in later \gaia\ data releases.
If $t$ is the timespan of observations, then whereas the parallax accuracy improves as $t^{1/2}$ on account of the increase in the number of observations, the proper motion accuracy improves as $t^{3/2}$ due to the additional increase in the observational baseline. This assumes systematic errors do not dominate, which is the case for most stars.
\gdr3\ is based on 34 months of data, whereas the final release, \gdr5, will be based on about 10.5 years (126 months) of data.
Using these factors, the parallax and proper motion accuracies should therefore increase by factors of 1.9 and 7.1 respectively
between \gdr3\ and \gdr5.
This suggests that while both the kinegeometric and geometric distance estimates should be more accurate with \gdr5\ data, the improvement for the kinegeometric ones might be larger.
  
I tested this expectation using the mock catalogue with similar simulations as before (section~\ref{sec:results_mock}), but now scaling down the parallax and proper motion uncertaintines by the above factors (1.9 and 7.1 respectively).
The results for the constant fraction sample are shown in the bottom block of Table~\ref{tab:results_mock_resid_statistics}. Looking first at the geometric distances -- which depend on the parallaxes but not the proper motions -- we see that the fractional bias is decreased from \gdr3\ to \gdr5\ by a factor of 1.8
and the fractional scatter is decreased by a factor of 1.35.
These improvements are not as large as the improvement in the parallaxes. This is expected, because the prior also plays a role:
The distance accuracies for the numerous stars with very low parallax SNRs ($\lesssim 1$) are only weakly dependent on their parallax SNRs because their distances are prior-dominated. For such stars, doubling the parallax SNR hardly changes their distance estimates and therefore hardly improves their accuracies.\footnote{Even with weaker priors, such as uniform in distance (or log distance), doubling the parallax SNR does not double the accuracy of the distance estimate, because of the nonlinear transformation between parallax and distance.}

For the kinegeometric distances, the fractional bias and scatter are decreased by factors of 1.55
and 1.35
respectively. Contrary to possible expectations, the kinegeometric distances do not improve more than the geometric distances: On average, at least, the larger improvement in the proper motions over the parallaxes has not helped.
The reason is that the proper motion only provides a distance estimate when combined with an assumed velocity distribution (the prior) via equation~\ref{eqn:transvel}.
Once the proper motions are sufficiently precise such that the width of this velocity distribution is the limiting factor in determining the distance precision, then improving the proper motions further will not help.
This is explained in appendix~\ref{sec:how_it_works}.
It appears that, on average at least, this limit has already been achieved in \gdr3.

\begin{figure*}
\begin{center}
\includegraphics[width=0.49\textwidth, angle=0]{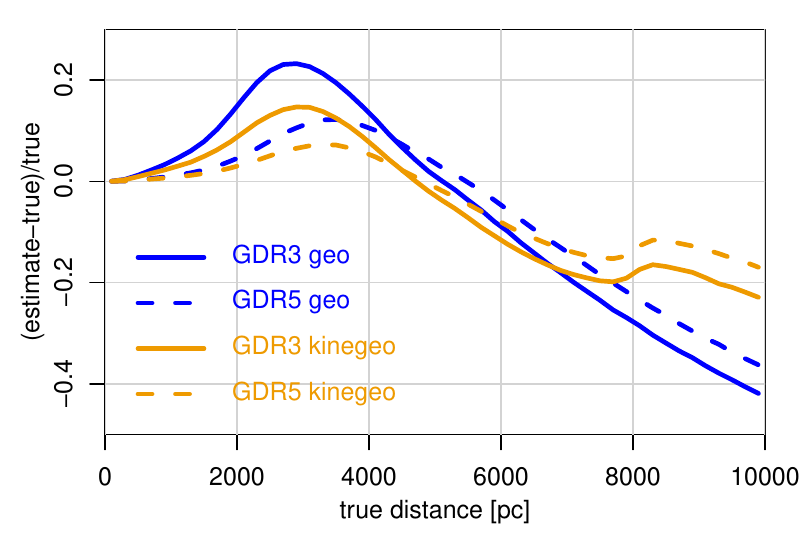}
\includegraphics[width=0.49\textwidth, angle=0]{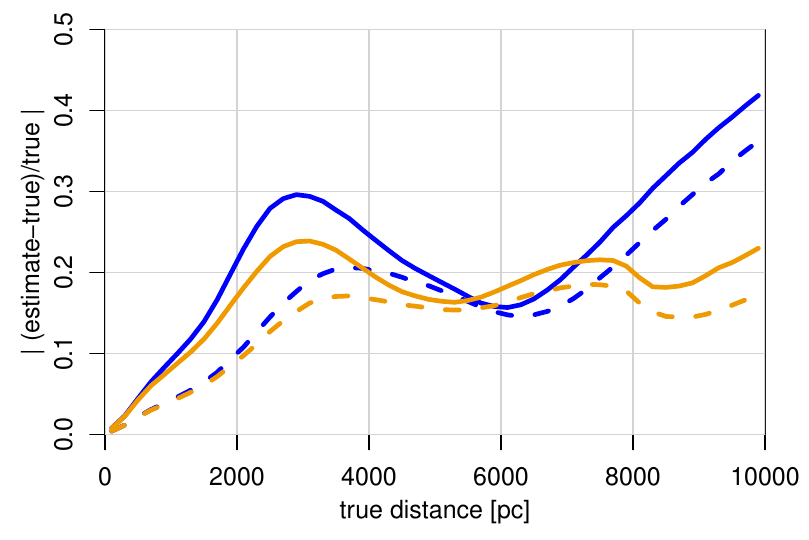}
  \caption{Median performance of the geometric (blue) and kinegeometric (orange) distance estimates as a function of distance for the constant fraction sample in the mock \gdr3\ (solid) and mock \gdr5\ (dashed) catalogues. The left panel shows the fractional bias in the estimates, the right panel the fractional scatter. The solid lines are the same as those in Figure~\ref{fig:results_mock_rMedKinogeo_bias_scatter_vs_distance}.
  \label{fig:results_mock_cfs_distance_bias_scatter_vs_distance_gdr3_gdr5}
}
\end{center}
\end{figure*}

The variation of the distance accuracy (bias and scatter) with distance for \gdr5\ is shown as the dashed lines in Figure~\ref{fig:results_mock_cfs_distance_bias_scatter_vs_distance_gdr3_gdr5}. The profiles are broadly similar to \gdr3\ (solid lines), but we see that the improvements in the accuracy vary with distance.
Below 2\,kpc\ the scatter in both the geometric and the kinegeometric distances is reduced by a factor of nearly two, whereas at around 5\,\kpc\ there is little improvement.

For the velocities the situation is similar (Table~\ref{tab:results_mock_resid_statistics}).
The scatter in the geometric velocities is expected to decrease by a factor of 1.35 times between \gdr3\ and \gdr5, 
and by a factor of 1.4 for kinegeometric velocities.
We might have expected much more improvement in both, because the proper motions improve by a factor of 7.1. But the mapping from proper motion to velocity depends also on the estimated distance, and the improvement in this, which is coming mostly from the parallaxes, is also a limiting factor in the improvement in the velocities.

I conclude that, in \gdr5, kinegeometric distances and velocities should be about 1.35 times more accurate than they are in \gdr3, when averaged over the entire \gaia\ sample in \gdr3.
These will still be only about 1.25 times more accurate that geometric distances and velocities.
The limited improvement is because the velocity priors are relatively broad.

While it may be worth using proper motions in \gdr5\ to improve the distance estimates in some regions, this comes at the cost of building in assumptions on velocities. If we are happy with this, then a better accuracy should be obtained by combining parallaxes and proper motions also with colours, magnitudes, and even spectra.

\section{Summary and conclusions}\label{sec:conclusions}

The kinegeometric method introduced in this paper estimates distances and transverse velocities together in a 3D posterior PDF using the parallaxes and proper motions. It uses a direction-dependence distance prior and a direction- and distance-dependent velocity prior. Geometric distances, in contrast, are obtained from just the parallaxes and the distance prior.

The performance was assessed on a random subset of the mock \gdr3\ catalogue.
This showed that on average, kinegeometric distances are 1.25 times more accurate than geometric ones (Table~\ref{tab:results_mock_resid_statistics}).
The median fractional distance residual (scatter) of the kinegeometric distances is 19\% averaged over all sources; averaged over the sky it is 11\%.
The main improvement brought by the proper motions is where the
parallax SNR is low ($\lesssim 4$; Figure~\ref{fig:results_mock_rMedKinogeo_bias_scatter_vs_parallaxSNR}).
There is a considerable spread in performance with position in the Galaxy, however.
Close to the Sun the kinegeometric distance errors are very small, then increase steadily to 20\% at 2\,\kpc\ and beyond (Figure~\ref{fig:results_mock_rMedKinogeo_bias_scatter_vs_distance}).
Kinegeometric distances have a low bias at high Galactic latitudes, but up to several percent in the bulge and disk, with complex patterns emerging at low latitudes due to the stellar density and prior variations (Figure~\ref{fig:results_mock_rMedGeo_rMedKinogeo_bias_scatter_skyplot}).
The biggest advantage of kinegeometric distances over geometric ones is for stars beyond about 7\,\kpc\ from the Sun.

The kinegeometric method also provides transverse velocity estimates. These can be compared to ``geometric'' velocities obtained by multiplying the geometric distances by proper motion. 
Both approaches have similar overall performance for stars within 7\,\kpc, but beyond that the kinegeometric velocities are considerably more accurate, including a lower bias (Figure~\ref{fig:results_mock_vtotMedKinogeo_bias_scatter_vs_distance}).

When applied to real \gdr3\ data we find that differences in the two distance estimates depend on direction and distance.
Beyond several \kpc, kinegeometric distances tend to be smaller on average (Figure~\ref{fig:results_cfs_kinogeo_minus_geo_over_geo_vs_geo}), which in particular is reflected by two regions in the bulge (Figure~\ref{fig:results_rMedGeo_rMedKinogeo_skyplot}). In the disk the picture is more complicated, a reflection of how the disk kinematics in the prior combine with the measured proper motions to inform the distance estimates.
The precisions (widths) of the posteriors are a statistically good estimate of the accuracies (Figure~\ref{fig:results_mock_06200_normalized_residual_distributions}), so we can use them to investigate the expected accuracy of the distance estimators in the absence of a ground truth.
The kinegeometric distances are more precise than the geometric ones for 62\% of stars, and (on average) over most of the sky (Figure~\ref{fig:results_cfs_distances_pwratio_skyplot}).

Overall, we see only a modest average improvement in distance accuracy when including proper motions.  This conclusion is based on the 34 months of \gaia\ data in \gdr3.  The SNR of the parallaxes and proper motions are expected to improve in future \gaia\ data releases, by factors of 1.9 and 7.1 respectively in the final release (\gdr5) assuming this is based on 10.5 years of data.  I find that this will improve the accuracy of the kinegeometric distances and velocities by
  a factor of 1.35.
  The geometric distances improve by a similar factor, meaning kinegeometric distances are still only about 1.25 times more accurate than geometric ones in \gdr5.
  The reason for this limited improvement is that
  proper motions only constrain distances when combined with an expected velocity, which we introduce as a distribution of velocities. If this prior is broad, then even very precise proper motions do not constrain the distances much (see appendix~\ref{sec:how_it_works}).
  This is very different from the situation with parallaxes, in which an increased accuracy of the parallax translates directly into an increased accuracy in the distance (provided the distance is not prior-dominated).
  Hence, for many stars in \gaia, the velocity distributions in the Galaxy are too broad for even high precision proper motions to provide strong constraints on distance.

\acknowledgments

I thank Morgan Fouesneau, Rene Andrae, and Sara Jamal for discussions on this project, and the referee for several useful suggestions.
This work was funded in part by the DLR (German space agency) via grant 50 QG 2102 and
has made use of data from the European Space Agency (ESA) mission \gaia\ (\url{http://www.cosmos.esa.int/gaia}), processed by the \gaia\ Data Processing and Analysis Consortium (DPAC, \url{http://www.cosmos.esa.int/web/gaia/dpac/consortium}). Funding for the DPAC has been provided by national institutions, in particular the institutions participating in the \gaia\ Multilateral Agreement. 
\facility{Gaia}

\appendix

\section{How the proper motion and velocity prior constrain distance}\label{sec:how_it_works}

The kinegeometric posterior of equation~\ref{eqn:kinegeopost} is a 3D distribution coupling distance and velocity. We can understand how the proper motion and velocity prior combine to produce distance estimates by considering a simpler version with just one velocity component and no parallax. In this case the posterior over distance is
\begin{equation}
  P(\dist \given \propm)
  \,=\, \int P(\dist, v \given \propm) \, dv
  \,=\, \int \frac{P(\propm \given \dist, v) P(\dist, v)}{P(\propm)} \,dv
  \,\propto \, P(\dist) \int P(\propm \given \dist, v) P(v \given \dist) \, dv \ .
  \label{eqn:kinegeopost2}
\end{equation}
We see that the proper motion likelihood constrains the distance only through the integral of its product with the velocity prior. This is quite different from the parallax, where $P(\dist \given \parallax) \propto P(\dist) P(\parallax \given \dist)$ without any integral.\footnote{This follows by replacing $\propm$ with $\parallax$ in equation~\ref{eqn:kinegeopost2} and then noting that the parallax likelihood is independent of $v$.}
Consequently, even if we measure a very precise proper motion, such that the likelihood $P(\propm \given \dist, v)$ for some $v$ in equation~\ref{eqn:kinegeopost2} is narrow, if the corresponding velocity prior $P(v \given \dist)$ is broader, the contribution to the posterior will also be broad.

\begin{figure}
  \begin{center}
\includegraphics[width=0.60\textwidth, angle=0]{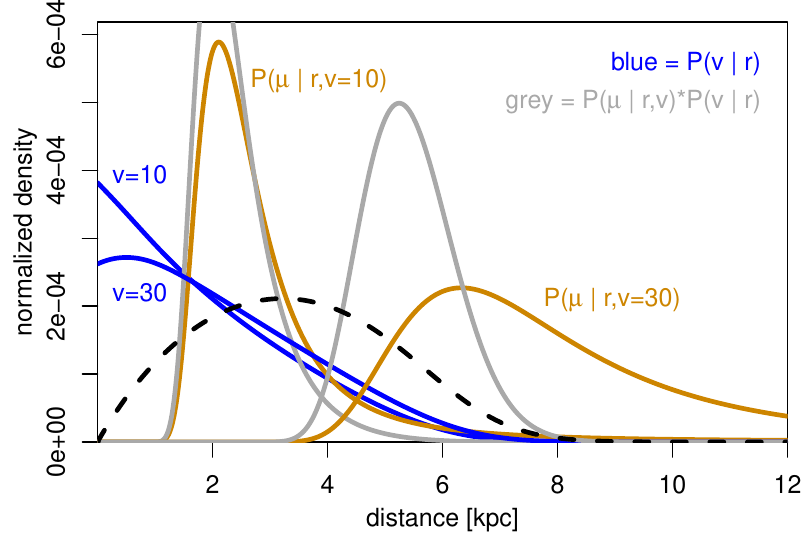}
\caption{Demonstration of how the proper motion likelihood ($P(\propm \given \dist, v)$, orange) combines with the velocity prior ($P(v \given \dist)$, blue) to constrain the distance via equation~\ref{eqn:kinegeopost2}, from a measured proper motion of $1 \pm 0.25$\,\maspyr.
This is shown for two different velocities, 10 and 30\,\kms.
The grey curve is the product of the orange and blue curves.
The black dashed line is the integral in equation~\ref{eqn:kinegeopost2} computed for a continuous range of velocities.
None of these functions are probability density functions in \dist, but all have been normalized to unit area over the range plotted.
\label{fig:demonstrate_prior_2_type2}
}
\end{center}
\end{figure}

This combination is demonstrated in Figure~\ref{fig:demonstrate_prior_2_type2}. I use the \vdec\ prior from HEALpixel 7593 (Figure~\ref{fig:velocity_prior_07593}) except that I take the absolute value of the mean of the velocity prior to more conveniently deal with positive values. The prior (blue curves) is shown for 10 and 30\,\kms. Both velocities show a preference for smaller distances.  The two orange curves are the likelihoods corresponding to these same two velocities assuming a proper motion measurement of $1 \pm 0.25$\,\maspyr.
1\,\maspyr\ and 10\,\kms\ corresponds to distance of 2.1\,\kpc\ (and 30\,\kms\ to  6.3\,\kpc; equation~\ref{eqn:transvel}), but  we see that this low SNR measurement does not constrain the distance much for either velocity.
Each grey curve is the product of the prior and likelihood, and therefore a term under the integral in equation~\ref{eqn:kinegeopost2} for a given velocity. 

\begin{figure}
  \begin{center}
    \includegraphics[width=0.49\textwidth, angle=0]{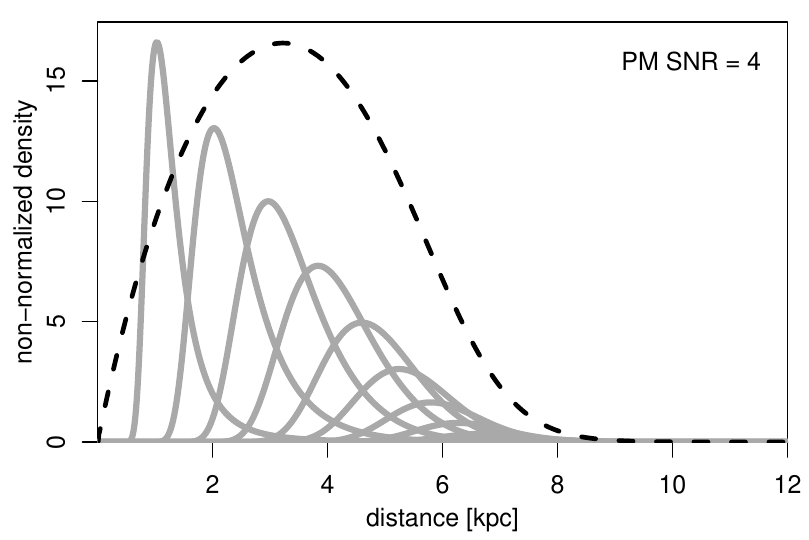}
    \includegraphics[width=0.49\textwidth, angle=0]{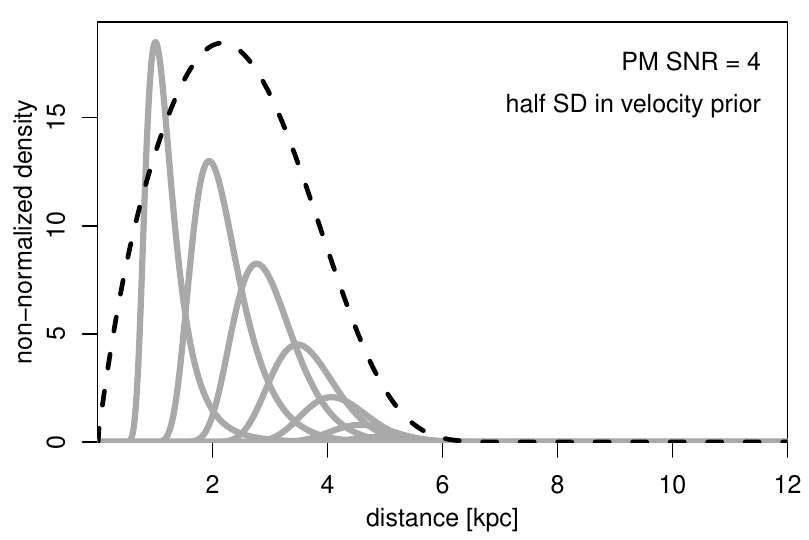}
    \includegraphics[width=0.49\textwidth, angle=0]{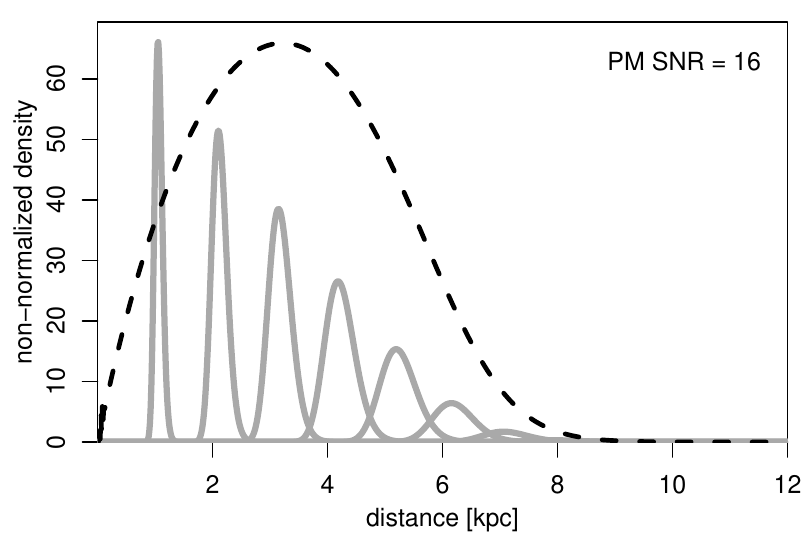}
    \includegraphics[width=0.49\textwidth, angle=0]{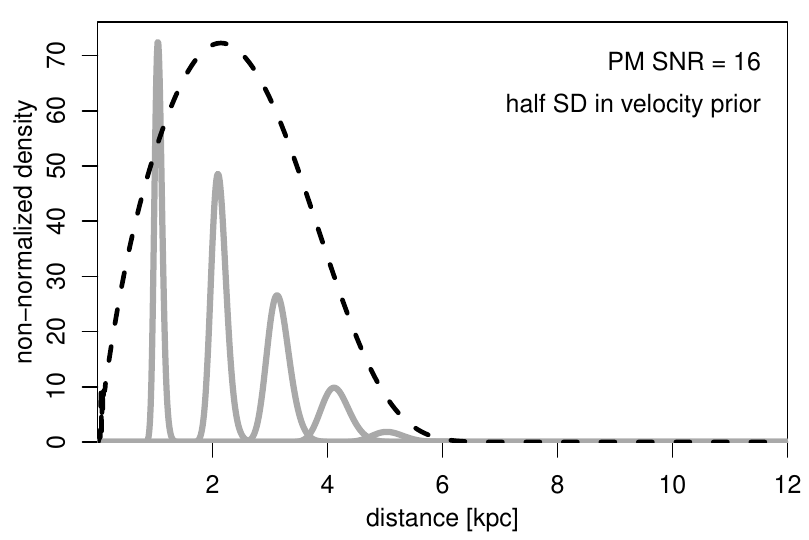}
    \caption{Demonstration of how the likelihood--prior product, $P(\propm \given \dist, v) P(v \given \dist)$, shown as the grey curves, sum to make the unnormalized distance posterior (black dashed line) via equation~\ref{eqn:kinegeopost2}. The grey curves are shown just for 10 velocities from 5 to 50\,\kms\ in steps of 5\,\kms,
but some are so flat as to be invisible. 
In all panels the measured proper motion is 1\,\maspyr. The top row is for a proper motion SNR of 4, the bottom for 16. The the right column if for a narrower prior with half the standard deviation of that in the left column.
None of the functions shown are probability density functions in \dist. The grey ones have been normalized to unit area over the range plotted. The black dashed curve has been scaled to have the same maximum as the largest grey peak.
\label{fig:demonstrate_prior_2_type1}
}
\end{center}
\end{figure}

If we repeat this for a continuous range of velocities, then the sum of all the resulting grey curves is
the integral in equation~\ref{eqn:kinegeopost2}, which is the unnormalized distance posterior (but omitting a distance prior).
This is illustrated in the top-left panel of 
Figure~\ref{fig:demonstrate_prior_2_type1}, where only a few of the grey curves with different velocities are shown, but the resulting integral (dashed black line) has been computed using a broad, dense grid of velocities.
We see from this plot that even though each contribution to the integral is relatively narrow, their sum, and therefore the posterior, is broad. Consequently, even if we have higher SNR proper motions -- and so narrower likelihoods and thus narrower grey distributions -- the integral is more or less unchanged. This is demonstrated in the bottom left panel of 
Figure~\ref{fig:demonstrate_prior_2_type1}, where the SNR is increased by a factor of four. This is the main point: improving the proper motion SNR does not necessarily improve the distance accuracy. This is why we expect little improvement in kinegeometric distances from \gdr3\ to \gdr5\ (section~\ref{sec:gdr5}).
If the prior is narrower, then we can get a narrower posterior for given data. This is illustrated in the right columns of Figure~\ref{fig:demonstrate_prior_2_type1}, where the standard deviation of the velocity prior has been halved at all distances. The proper motion likelihoods are unchanged (compared to the panel on the left), but some are now more suppressed by the prior.


\section{Random errors cause bias}\label{sec:cause_of_bias}

One might think that, because the priors are built from the mock catalogue, there should be no bias in the results on the mock catalogue in section~\ref{sec:results_mock} (for either distance estimate). This is incorrect for at least three reasons. First, the mock catalogue used in the performance assessment is noisy. Although the noise is symmetric (Gaussian) in the parallax, it is not symmetric in the distance due to their nonlinear relation ($\dist \sim 1/\parallax$; see \citealt{2015PASP..127..994B}).
Second, the distance and velocity priors are fits, so do not represent all the variances in the mock data. We may in fact not want a prior tuned exactly to all the details and specific choices of the mock catalogue, because this would be a rather strong prior.
Third, in reflecting the true Galaxy, the distance prior has the bulk of its probability mass within a few kpc.
Stars beyond these distances tend to have low parallax SNRs, i.e.\ the data are relatively uninformative, so the prior will tend to pull the inferred distance below the true distance for these stars.
This is why we see larger biases at larger distances in Figure~\ref{fig:results_mock_rMedKinogeo_bias_scatter_vs_distance}.
Nearby low parallax SNR stars will tend to have their distances pulled up by the prior for the same reason, but this is much less common because there are fewer nearby stars, and because nearby stars tend to have higher parallax SNRs on account of their larger parallaxes and brighter apparent magnitudes.


``Not using a prior'' is not a solution to this problem, because a prior is invariably present implicitly.
Indeed, if we just invert a parallax to estimate the distance, the biases are much larger \citep{2015PASP..127..994B}.
A prior uniform in distance or velocity is a poor choice, and much more biased than the present choice: A uniform distance prior is equivalent to assuming that the stellar number density drops as $1/r^2$ from the Sun,
which is demonstrably wrong. A uniform velocity distribution relative to the Sun contradicts the existence of the different components of the Galaxy.
Unless one is prepared to use highly tuned priors, low SNR data give low SNR inferences.
The only other solution is to acquire better data.


\bibliographystyle{aasjournal}
\bibliography{distances,gaia}

\end{document}